\documentclass[12pt]{article}
\usepackage[a4paper,margin=1in,includefoot,footskip=50pt]{geometry}
\usepackage[utf8]{inputenc}
\usepackage[T1]{fontenc}
\usepackage{lmodern}
\usepackage[english]{babel}
\DeclareUnicodeCharacter{03B2}{\ensuremath{\beta}}
\DeclareUnicodeCharacter{03B3}{\ensuremath{\gamma}}
\usepackage{microtype}
\usepackage{graphicx}
\usepackage{booktabs}
\usepackage{array}
\usepackage{tabularx}
\usepackage{adjustbox}
\usepackage{caption}
\usepackage{placeins}
\usepackage{mathtools}
\usepackage{amssymb}
\usepackage{xcolor}
\usepackage{xurl}
\usepackage{tikz}
\usetikzlibrary{arrows.meta,positioning,shapes.geometric}
\usepackage[numbers,sort&compress]{natbib}
\usepackage[colorlinks=true,linkcolor=blue,citecolor=blue,urlcolor=blue]{hyperref}
\usepackage{fancyhdr}

\fancypagestyle{firstpage}{
  \fancyhf{}

  \fancyfoot[L]{%
    \begin{minipage}[t]{0.90\textwidth}
    \footnotesize
    \textsuperscript{*}Corresponding author: Mohammad Alamgir Chowdhury\\
    Department of Statistics, University of Manitoba, Winnipeg, Canada\\
    Email: \href{mailto:chowdh62@myumanitoba.ca}{chowdh62@myumanitoba.ca}
    \end{minipage}
  }
  \fancyfoot[R]{\footnotesize\thepage}
}
\title{Connectome-Based Modeling of Mutation-Specific
Amyloid-\texorpdfstring{$\beta$}{beta} Aggregation in Familial Alzheimer's Disease}

\author{
Mohammad Alamgir Chowdhury$^{a,*}$,
Hina Shaheen$^{b}$\\[0.5em]
\small $^{a}$ Department of Statistics,
University of Manitoba, Winnipeg, Canada\\
\small $^{b}$Faculty of Science, Department of Statistics, 
University of Manitoba, Winnipeg, Canada
}

\date{}

\hypersetup{
  pdftitle={Connectome-Based Modeling of Mutation-Specific Amyloid-beta Aggregation in Familial Alzheimer's Disease},
  pdfauthor={Mohammad Alamgir Chowdhury and Hina Shaheen}
}

\begin{document}
\maketitle
\thispagestyle{firstpage}

\begin{abstract}
Familial amyloid-$\beta$ (A$\beta$) variants can alter aggregation kinetics,
but how these molecular differences interact with structural brain connectivity
remains incompletely understood. We developed a mutation-aware mechanistic
framework coupling a coarse-grained monomer--oligomer--fibril
aggregation--fragmentation system to graph diffusion on the 540-node largest
connected component of the Budapest Reference Connectome. Experimental
A$\beta_{42}$ nucleation scores were mapped exponentially to relative primary
nucleation rates for seven variants, with wild type as the reference.
Robustness was evaluated through mutation-score uncertainty, alternative kinetic
mappings, global parameter sensitivity, seed and edge-weight perturbations,
20 degree-preserving randomized connectomes, spatial propagation analysis,
synthetic ABC-SMC parameter recovery, posterior prediction, and reaction-level
Chemical Langevin simulations. E22G exhibited the earliest threshold crossing
and greatest cumulative oligomer burden, whereas A2V was delayed under the
selected mapping. Mutation rankings persisted across the tested network
perturbations, although regional burden patterns changed with topology.
Connectome distance from the seed regions was strongly associated with later
oligomer arrival (Spearman $\rho\approx0.92$). Under inferred parameter
uncertainty, the mean timing-rank correlation was 0.990 and cumulative-burden
ordering was preserved in every posterior draw; exact peak-amplitude ordering
was less stable. Stochastic ensemble medians retained the deterministic mutation
ordering despite overlap among individual trajectories. Within this
proof-of-concept framework, mutation-dependent kinetics primarily influence
relative aggregation timing and cumulative burden, while structural connectivity
shapes spatial propagation. Nominal background rates, model-time units, and
synthetic parameter recovery limit interpretation to mechanistic comparisons
rather than clinically calibrated prediction.
\end{abstract}

\noindent\textbf{Keywords:} amyloid-$\beta$; familial Alzheimer's disease; structural connectome; aggregation--fragmentation; parameter uncertainty; stochastic modelling.

\section{Introduction}
\label{SE1}

Alzheimer's disease (AD) is a neurodegenerative disorder that occurs predominantly in sporadic form, although a smaller proportion of cases arise through autosomal-dominant inheritance associated with pathogenic variants in genes such as amyloid precursor protein (\textit{APP}) and the presenilins \cite{Dunne2016,Bekris2010,Bateman2011,Nasica2015}. Mutations within or near the \textit{APP} region encoding amyloid-$\beta$ (A$\beta$) can alter peptide charge, hydrophobicity, conformation, and aggregation kinetics, thereby influencing the formation of soluble oligomers and fibrillar deposits that are central features of AD pathology \cite{Kim2008,Li2019}. A key early step in A$\beta$ aggregation is nucleation, in which soluble peptides form a stable nucleus that permits subsequent aggregate growth. Under physiological conditions, this initial step is often kinetically slow and energetically unfavorable \cite{Buell2017}. Once nucleation occurs, additional processes including elongation, secondary nucleation, fragmentation, and conversion can accelerate aggregate formation and amplify oligomer and fibril burden \cite{Baldwin2011,Knowles2014}. Mutations that alter nucleation behavior may therefore produce substantial differences in downstream aggregation dynamics \cite{Chiti2017}.

Familial AD mutations provide a useful setting in which to study how single-amino-acid substitutions affect A$\beta$ aggregation \cite{Estrada-Rodriguez2019}. The present study considers the variants A2V, A21G, D23N, E22G, E22Q, E22K, and H6R, with wild type (WT) used as the reference condition \cite{Davidson2020}. These mutations have been associated with distinct aggregation and neuropathological phenotypes. E22G (Arctic) has been linked to enhanced protofibril formation and accelerated aggregation \cite{Norlin2012}; E22Q (Dutch), E22K (Italian), and D23N (Iowa) have been associated with cerebral amyloid angiopathy and altered amyloid deposition \cite{Levy1990,Melchor2000,Mok2014}; A21G (Flemish) has been linked to altered APP processing and A$\beta$ generation \cite{Tang2014}; H6R (English) has been associated with altered early oligomerization dynamics \cite{Viet2014}; and A2V has been reported to modify A$\beta$ aggregation propensity \cite{Zheng2015}. These mutation-specific differences motivate a modeling framework in which aggregation kinetics are not assumed to be identical across variants. Recent high-throughput mutational measurements provide quantitative mutation scores that can be used to represent relative effects on A$\beta$ nucleation \cite{Faure2024}. In the present study, these scores are used to scale the primary nucleation rate in the main model, while alternative mappings involving secondary nucleation are examined as mechanism-uncertainty analyses. This provides a direct link between experimentally informed mutation effects and a coarse-grained aggregation model.

To connect molecular aggregation kinetics with large-scale brain organization, we developed a mutation-aware mechanistic model on the Budapest Reference Connectome \cite{szalkai2017budapest}. Local A$\beta$ dynamics are represented by a coarse-grained Smoluchowski-type aggregation--fragmentation system containing three effective species: monomers, oligomers, and fibrils. The reaction system includes primary nucleation, secondary nucleation, elongation, conversion, fragmentation, and clearance. These local dynamics are coupled across brain regions through diffusion on the structural connectome Laplacian, producing a network aggregation--fragmentation--diffusion model. The central aim of the study is to determine whether experimentally informed mutation-specific aggregation kinetics produce robust differences in global amyloid dynamics and how those differences interact with structural network topology. The analysis therefore distinguishes between two related questions: whether mutation-specific kinetics determine the timing and cumulative burden of aggregation, and whether the structural connectome determines the regional pattern through which that burden is expressed.

To assess robustness, the model is examined under several complementary analyses. Mutation-score uncertainty is propagated through the aggregation model, and global sensitivity analysis is used to identify parameters that most strongly influence aggregation timing, burden, and spatial spread. Robustness is further evaluated across alternative seed configurations, connectome weighting schemes, and degree-preserving randomized connectomes. Spatial propagation is assessed by relating regional activation times and burden to network distance from the seed set. A likelihood-free ABC-SMC analysis is used to examine internal parameter recoverability, followed by posterior-predictive propagation of parameter uncertainty through the full connectome. Finally, a Chemical Langevin formulation is used to quantify reaction-level stochastic variability around the deterministic dynamics, while interregional network transport remains deterministic.

The purpose of this study is to develop and test a mechanistic proof-of-concept framework linking mutation-specific A$\beta$ aggregation kinetics to connectome-constrained brain-wide propagation. The resulting model is intended to clarify which aspects of the dynamics are primarily controlled by molecular kinetics and which are shaped by large-scale structural connectivity.

\section{Modeling and Computational Methods}
\label{sec:methods}

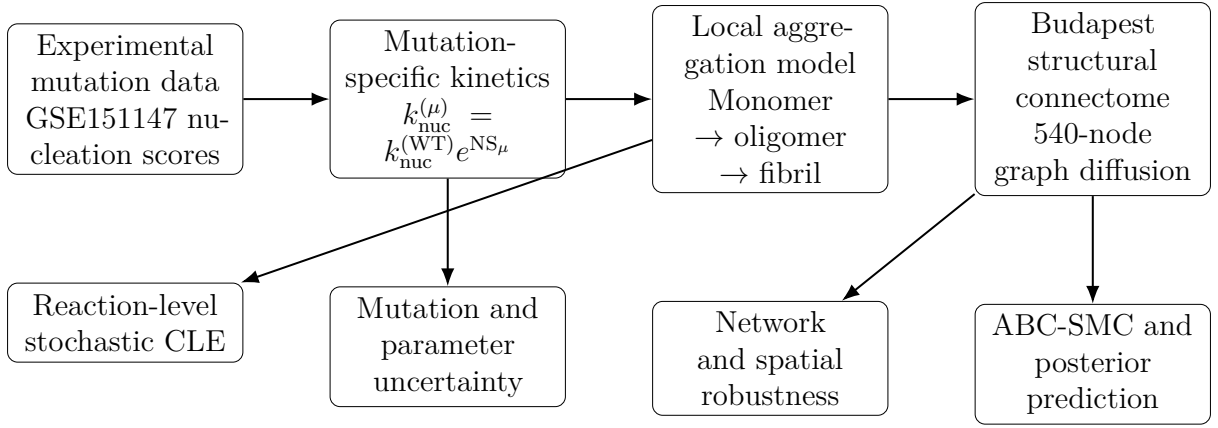
\begin{figure}[htbp]
\centering
\begin{adjustbox}{max width=\textwidth}
\begin{tikzpicture}[
    node distance=1.3cm and 1.2cm,
    box/.style={
        rectangle,
        rounded corners,
        draw,
        align=center,
        minimum width=3.2cm,
        minimum height=1.1cm,
        text width=3.0cm
    },
    arrow/.style={
        -{Latex[length=2.5mm]},
        thick
    }
]

\node[box] (data) {
Experimental mutation data\\
GSE151147 nucleation scores
};

\node[box, right=of data] (mutation) {
Mutation-specific kinetics\\
$k_{\mathrm{nuc}}^{(\mu)}
=
k_{\mathrm{nuc}}^{(\mathrm{WT})}
e^{\mathrm{NS}_{\mu}}$
};

\node[box, right=of mutation] (reaction) {
Local aggregation model\\
Monomer $\rightarrow$ oligomer $\rightarrow$ fibril
};

\node[box, right=of reaction] (network) {
Budapest structural connectome\\
540-node graph diffusion
};

\draw[arrow] (data) -- (mutation);
\draw[arrow] (mutation) -- (reaction);
\draw[arrow] (reaction) -- (network);

\node[box, below=1.5cm of mutation] (uncertainty) {
Mutation and parameter\\
uncertainty
};

\node[box, below=1.5cm of reaction] (robustness) {
Network and spatial\\
robustness
};

\node[box, below=1.5cm of network] (abc) {
ABC-SMC and\\
posterior prediction
};

\node[box, below=1.5cm of data] (stochastic) {
Reaction-level\\
stochastic CLE
};

\draw[arrow] (mutation) -- (uncertainty);
\draw[arrow] (network) -- (robustness);
\draw[arrow] (network) -- (abc);
\draw[arrow] (reaction) -- (stochastic);

\end{tikzpicture}
\end{adjustbox}

\caption{Overview of the mutation-aware A$\beta$ modelling framework.
Experimentally derived nucleation scores were mapped to
mutation-specific primary nucleation rates and incorporated into a
coarse-grained aggregation--fragmentation model. Local reaction dynamics
were coupled across the Budapest Reference Connectome using graph
diffusion. Robustness was examined through mutation and parameter
uncertainty, seed and network perturbations, spatial propagation
analyses, synthetic ABC-SMC parameter recovery, posterior-predictive
simulations, and reaction-level stochastic simulations.}
\label{fig:model_workflow}
\end{figure}

The modelling framework linked experimentally derived mutation-specific
A$\beta_{42}$ nucleation information with a coarse-grained
aggregation--fragmentation model and a structural brain connectome. The
main analysis consisted of a deterministic mutation-aware network model,
followed by uncertainty, sensitivity, network-robustness, synthetic
parameter-recovery, posterior-predictive, and stochastic analyses. An
overview of the complete modelling workflow is shown in
Figure~\ref{fig:model_workflow}.

Mutation-specific A$\beta_{42}$ nucleation scores were obtained from
the GSE151147 deep-mutational scanning dataset \cite{Seuma2021}. These
scores describe relative changes in nucleation compared with wild type
(WT) and are therefore treated as experimental phenotypes rather than
direct measurements of microscopic kinetic rate constants. Seven
familial Alzheimer disease variants were considered: A2V, H6R, A21G,
E22K, D23N, E22Q, and E22G, with WT used as the reference condition.
For mutation $\mu$, the nucleation score is denoted by
$\mathrm{NS}_{\mu}$, with $\mathrm{NS}_{\mathrm{WT}}=0$.

To incorporate the experimental mutation information into the
mechanistic model, the nucleation score was converted to a positive
multiplicative factor,

\begin{equation}
f_{\mu}
=
\exp\!\left(\mathrm{NS}_{\mu}\right),
\label{eq:mutation_scale}
\end{equation}

and the mutation-specific primary nucleation rate was defined as

\begin{equation}
k_{\mathrm{nuc}}^{(\mu)}
=
k_{\mathrm{nuc}}^{(\mathrm{WT})}
\exp\!\left(\mathrm{NS}_{\mu}\right).
\label{eq:mutation_knuc}
\end{equation}

This exponential transformation is an operational modelling choice
that converts the relative nucleation phenotype into a relative
kinetic effect. The score itself is not interpreted as a free-energy
barrier or as a directly measured rate constant. In the primary
network model, mutation dependence was applied only to
$k_{\mathrm{nuc}}$, while $k_{\mathrm{sec}}$ remained at its WT value.
Alternative mappings in which the mutation score affected secondary
nucleation alone or both primary and secondary nucleation were examined
separately as mechanism-uncertainty analyses. The processed GSE247583
dataset was retained only as an auxiliary dataset for comparison and
quality checking and was not used to define the primary
mutation-specific parameterization.

Large-scale structural connectivity was represented using the Budapest
Reference Connectome v3.0 \cite{szalkai2017budapest}. The original
GraphML network contained 1015 nodes and 1298 weighted undirected
edges. Because the complete graph contained disconnected components,
all propagation analyses were performed on its largest connected
component, giving a connected network of 540 nodes and 1296 edges.
Structural connection strength was represented using the Budapest
connectome attribute
\texttt{number\_of\_fiber\_per\_fiber\_length\_mean}. Let
$A^{\mathrm{raw}}$ denote the corresponding weighted adjacency
matrix. The raw strength of node $i$ was
$s_i^{\mathrm{raw}}=\sum_{j=1}^{N}A_{ij}^{\mathrm{raw}}$. To place
the structural weights on a common numerical scale, the adjacency
matrix was divided by the maximum empirical raw node strength,
$\kappa=\max_i s_i^{\mathrm{raw}}=782$, giving
$A=A^{\mathrm{raw}}/\kappa$.

The weighted-degree matrix was defined by
$D_{ii}=\sum_{j=1}^{N}A_{ij}$, and the corresponding combinatorial
graph Laplacian was

\begin{equation}
L=D-A.
\label{eq:laplacian}
\end{equation}

\begin{figure}[htbp]
    \centering
    \includegraphics[width=0.8\textwidth]{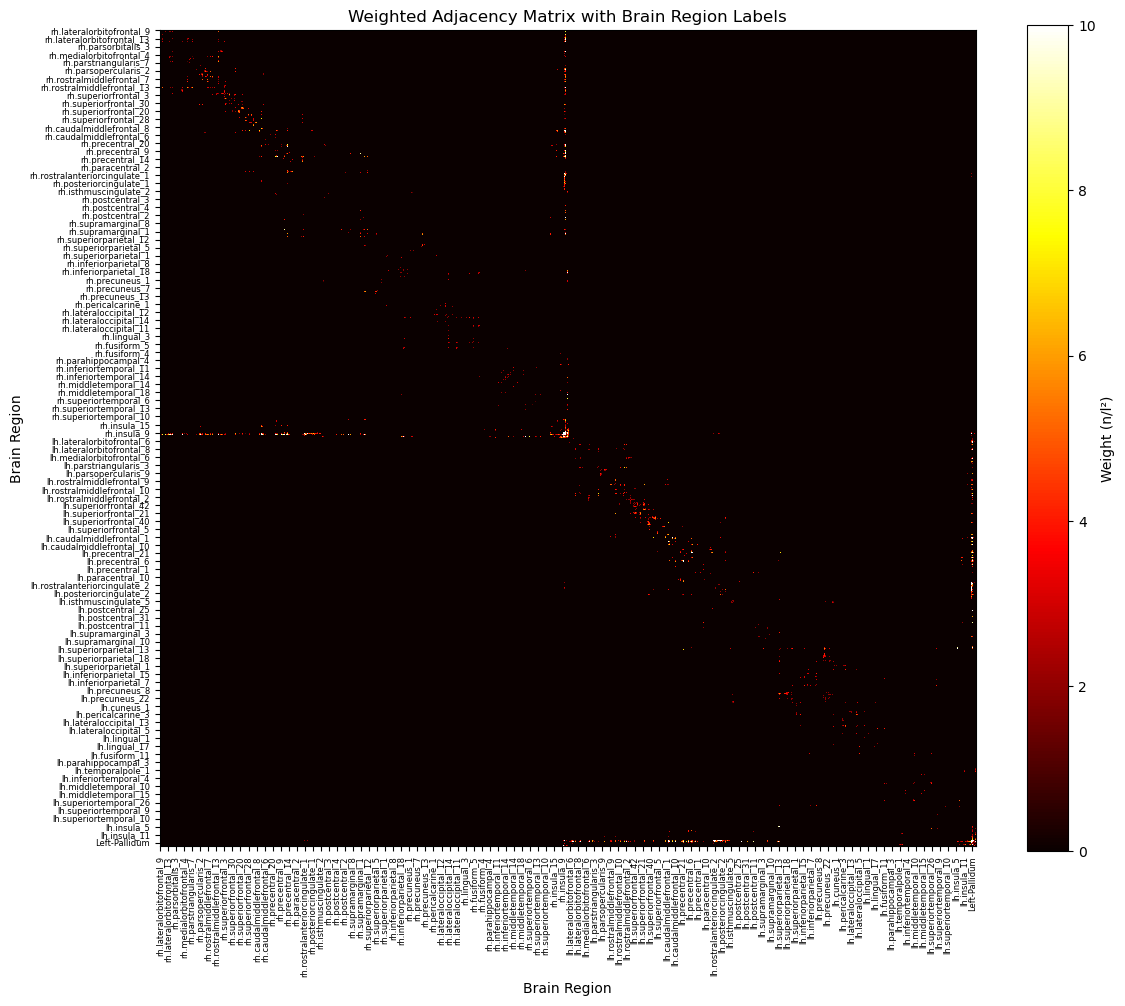}
    \caption{Weighted adjacency matrix of the 540-node Budapest
    Reference Connectome used in the propagation model. Matrix entries
    represent normalized structural connection strengths between brain
    regions. The adjacency matrix was used to construct the graph
    Laplacian governing interregional A$\beta$ diffusion.}
    \label{fig:adjacency_matrix}
\end{figure}

The negative Laplacian, $-L$, was used to model diffusion between
structurally connected brain regions, consistent with established
network-diffusion formulations \cite{raj2012network,Shaheen2023}.
The weighted adjacency matrix used to construct the Laplacian is shown
in Figure~\ref{fig:adjacency_matrix}.

Posterior cingulate and precuneus regions were selected as candidate
seed locations because these regions are associated with early
cortical A$\beta$ accumulation \cite{Palmqvist2017}. Thirty-six
candidate nodes were identified from their anatomical labels, and the
18 nodes with the highest raw structural strengths were used as the
primary seed set $\mathcal{S}$. The corresponding seed indicator was

\begin{equation}
\chi_i=
\begin{cases}
1, & i\in\mathcal{S},\\
0, & i\notin\mathcal{S}.
\end{cases}
\label{eq:seed_indicator}
\end{equation}

Initial monomer and oligomer concentrations were
$c_{1,i}(0)=1$ and $c_{2,i}(0)=0$ at all nodes. A small fibrillar seed
was introduced only at the seed locations,

\begin{equation}
c_{3,i}(0)=
\begin{cases}
10^{-6}, & i\in\mathcal{S},\\
0, & i\notin\mathcal{S}.
\end{cases}
\label{eq:initial_conditions}
\end{equation}

The value $10^{-6}$ was used only to initiate the model and should not
be interpreted as a measured regional A$\beta$ concentration.

\subsection{Reaction network and deterministic connectome model}
\label{sec:deterministic_model}

At each brain region, the model tracks three effective A$\beta$
species: monomers $c_1$, oligomers $c_2$, and fibrils $c_3$. The local
reaction system is a coarse-grained Smoluchowski-type
aggregation--fragmentation model rather than a full aggregate-size
distribution. It includes monomer production, primary nucleation,
secondary nucleation, elongation, oligomer-to-fibril conversion,
fibril fragmentation, and species-specific clearance
\cite{Cohen2013,Knowles2014,Shaheen2023}. This reduced representation
retains the main reaction pathways while remaining computationally
feasible for repeated simulation on the 540-node structural
connectome.

\begin{figure}[htbp]
    \centering
    \includegraphics[width=0.72\textwidth]{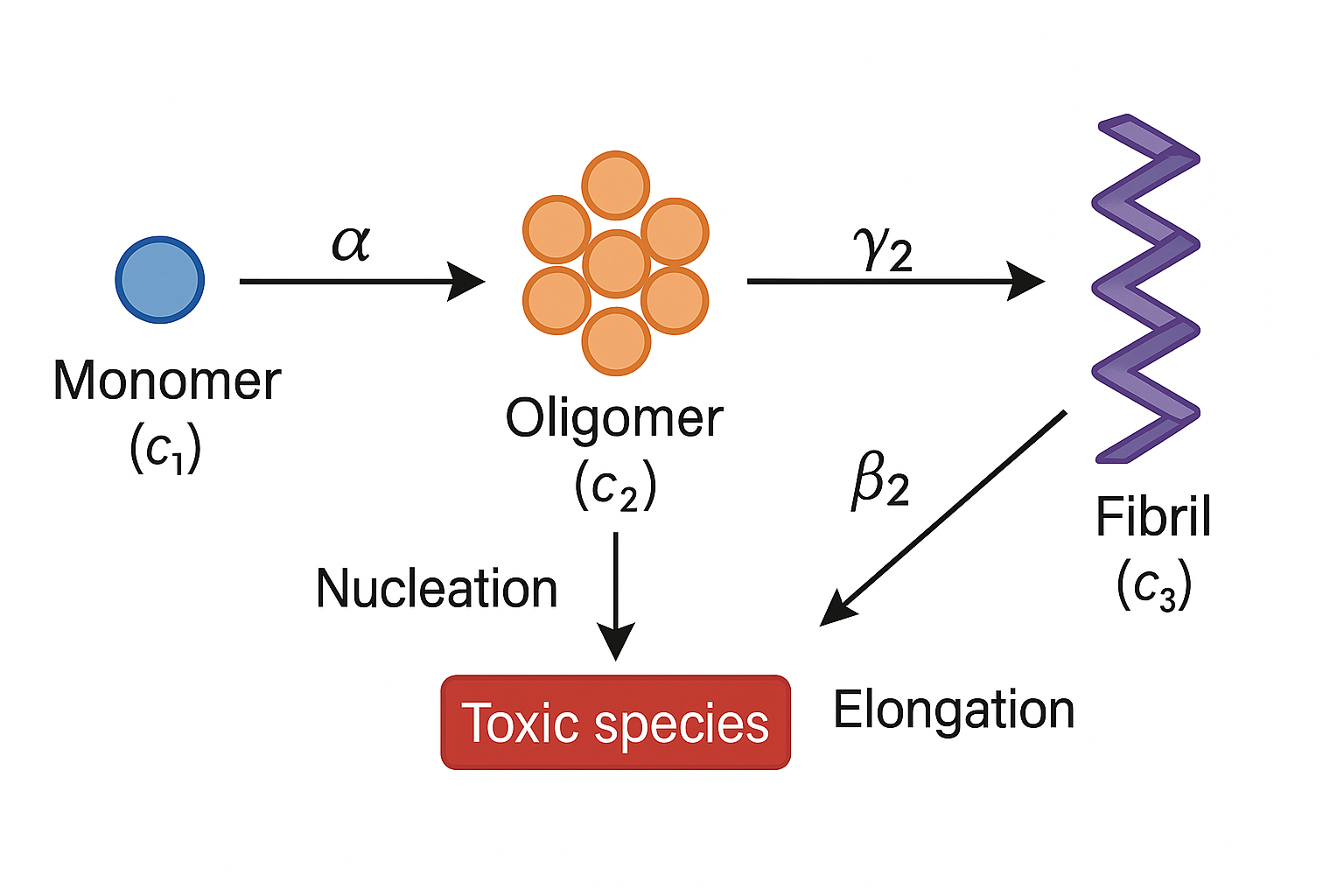}
    \caption{Schematic representation of the coarse-grained A$\beta$
    aggregation model. Monomers ($c_1$) form oligomers ($c_2$) through
    primary and secondary nucleation. Oligomers are converted to
    fibrils ($c_3$), while fibrils grow through elongation and generate
    additional oligomeric material through fragmentation. In the
    primary mutation-aware model, experimentally derived nucleation
    scores scale the primary nucleation rate $k_{\mathrm{nuc}}$.
    Alternative effects on secondary nucleation were examined
    separately in the mechanism-uncertainty analysis.}
    \label{fig:kinetic_schematic}
\end{figure}

The local reaction channels, propensities, and corresponding
stoichiometric changes are summarized in
Table~\ref{tab:reaction_channels}. Primary nucleation is restricted to
the seed regions through $\chi_i$, whereas secondary nucleation,
elongation, conversion, fragmentation, production, and clearance are
evaluated throughout the network.

\begin{table}[htbp]
\centering
\small
\caption{Reaction channels in the coarse-grained A$\beta$ aggregation model.}
\label{tab:reaction_channels}
\renewcommand{\arraystretch}{1.25}

\begin{adjustbox}{max width=\textwidth}
\begin{tabular}{
p{3.0cm}
p{3.4cm}
p{5.0cm}
p{3.8cm}
}
\toprule
\textbf{Process} &
\textbf{Reaction} &
\textbf{Propensity} &
\textbf{Stoichiometric change} \\
\midrule

Monomer production &
$\emptyset \rightarrow c_1$ &
$a_i^{(\mathrm{src})}=s$ &
$(+1,0,0)$ \\[3pt]

Primary nucleation &
$n c_1 \rightarrow \alpha_{\mathrm{nuc}}c_2$ &
$a_i^{(\mathrm{nuc})}
=
\chi_i k_{\mathrm{nuc}}^{(\mu)}c_{1,i}^{\,n}$ &
$(-n,+\alpha_{\mathrm{nuc}},0)$ \\[3pt]

Secondary nucleation &
$m c_1+c_3 \rightarrow \alpha_{\mathrm{sec}}c_2+c_3$ &
$a_i^{(\mathrm{sec})}
=
k_{\mathrm{sec}}c_{1,i}^{\,m}c_{3,i}^{\,p}$ &
$(-m,+\alpha_{\mathrm{sec}},0)$ \\[3pt]

Elongation &
$c_1+c_3 \rightarrow (1+\beta_{\mathrm{elong}})c_3$ &
$a_i^{(\mathrm{elong})}
=
k_{\mathrm{elong}}c_{1,i}c_{3,i}$ &
$(-1,0,+\beta_{\mathrm{elong}})$ \\[3pt]

Conversion &
$c_2\rightarrow c_3$ &
$a_i^{(\mathrm{conv})}
=
k_{\mathrm{conv}}c_{2,i}$ &
$(0,-1,+1)$ \\[3pt]

Fragmentation &
$c_3\rightarrow c_2$ &
$a_i^{(\mathrm{frag})}
=
k_{\mathrm{frag}}c_{3,i}$ &
$(0,+1,-1)$ \\[3pt]

Clearance &
$c_x\rightarrow\emptyset$ &
$a_i^{(\mathrm{clr},x)}
=
\delta_xc_{x,i}$ &
$-\mathbf e_x,\quad x\in\{1,2,3\}$ \\
\bottomrule
\end{tabular}
\end{adjustbox}
\end{table}

Let $\mathcal{R}$ denote the set of local reaction channels and
$\Delta_{xr}$ the stoichiometric change in species $x$ caused by
reaction $r$. The complete deterministic dynamics at node $i$ can be
written compactly as

\begin{equation}
\frac{dc_{x,i}}{dt}
=
\sum_{r\in\mathcal R}
\Delta_{xr}a_{r,i}(\mathbf c)
-
\rho d_x(Lc_x)_i,
\label{eq:general_network}
\end{equation}

where the first term represents local aggregation reactions and the
second term represents graph diffusion through the structural
connectome. 

The parameter $\rho$ controls the overall strength of
network transport and $d_x$ controls the relative mobility of species
$x$. The relative diffusion coefficients were
$d_1=1$, $d_2=2^{-1/3}$, and $d_3=3^{-1/3}$, with nominal
network-diffusion scale $\rho=0.05$.

Using the reaction propensities in
Table~\ref{tab:reaction_channels}, the three-species deterministic
network equations are

\begin{align}
\frac{dc_{1,i}}{dt}
&=
s-\delta_1c_{1,i}
-n\,a_i^{(\mathrm{nuc})}
-m\,a_i^{(\mathrm{sec})}
-a_i^{(\mathrm{elong})}
-\rho d_1(Lc_1)_i,
\label{eq:c1_network}
\\[5pt]
\frac{dc_{2,i}}{dt}
&=
\alpha_{\mathrm{nuc}}a_i^{(\mathrm{nuc})}
+
\alpha_{\mathrm{sec}}a_i^{(\mathrm{sec})}
+
a_i^{(\mathrm{frag})}
-
a_i^{(\mathrm{conv})}
-
\delta_2c_{2,i}
-
\rho d_2(Lc_2)_i,
\label{eq:c2_network}
\\[5pt]
\frac{dc_{3,i}}{dt}
&=
\beta_{\mathrm{elong}}a_i^{(\mathrm{elong})}
+
a_i^{(\mathrm{conv})}
-
a_i^{(\mathrm{frag})}
-
\delta_3c_{3,i}
-
\rho d_3(Lc_3)_i.
\label{eq:c3_network}
\end{align}

Because $L=D-A$, the negative sign before the Laplacian corresponds to
ordinary diffusion from regions of higher concentration toward
structurally connected regions of lower concentration.

Because several effective rate constants in the present coarse-grained
three-species formulation are not directly identifiable from available
\textit{in vivo} data, the baseline kinetic, clearance, production, and
network-transport parameters were treated as nominal reference values rather
than as experimentally measured biological constants. This approach is
consistent with previous qualitative Smoluchowski-type models of amyloid
aggregation and propagation, in which reference parameter regimes were used
to investigate mechanistic behaviour rather than to make quantitatively
calibrated clinical predictions
\cite{Achdou2013,Bertsch2017,Fornari2020}.

The nominal values were therefore selected to define a numerically stable
reference regime in which primary nucleation, secondary nucleation,
elongation, conversion, fragmentation, clearance, and network transport
could all contribute to the simulated dynamics.

Accordingly, interpretation
of the model focuses on relative mutation-dependent differences, parameter
sensitivity, and robustness of the resulting ordering rather than on the
absolute simulated concentrations or times. To assess dependence on this
reference parameterization, the principal kinetic, clearance, production,
and diffusion parameters were subsequently varied simultaneously over
$0.1$--$10$ times their respective nominal values using Latin hypercube
sampling in the global sensitivity analysis. The nominal parameter values used in the proof-of-concept simulations are
summarized in Table~\ref{tab:nominal_parameters}.

\begin{table}[htbp]
\centering
\small
\caption{Nominal reference parameters of the deterministic aggregation model. These values define the baseline proof-of-concept parameterization and are not interpreted as experimentally measured \textit{in vivo} kinetic constants.}
\label{tab:nominal_parameters}
\renewcommand{\arraystretch}{1.15}

\begin{tabular}{lll}
\toprule
\textbf{Parameter} &
\textbf{Nominal value} &
\textbf{Role} \\
\midrule

$k_{\mathrm{nuc}}^{(\mathrm{WT})}$ & $1.0\times10^{-3}$ & Primary nucleation \\
$k_{\mathrm{sec}}$                  & $1.0\times10^{-3}$ & Secondary nucleation \\
$k_{\mathrm{elong}}$                & $1.0\times10^{-1}$ & Fibril elongation \\
$k_{\mathrm{frag}}$                 & $1.0\times10^{-4}$ & Fibril fragmentation \\
$k_{\mathrm{conv}}$                 & $1.0\times10^{-1}$ & Oligomer-to-fibril conversion \\
$\delta_1$                           & $5.0\times10^{-3}$ & Monomer clearance \\
$\delta_2$                           & $2.0\times10^{-3}$ & Oligomer clearance \\
$\delta_3$                           & $1.0\times10^{-3}$ & Fibril clearance \\
$s$                                  & $1.0\times10^{-2}$ & Monomer production \\
$n$                                  & $2$                 & Primary nucleation order \\
$m$                                  & $1$                 & Secondary nucleation monomer order \\
$p$                                  & $1$                 & Secondary nucleation fibril order \\
$\alpha_{\mathrm{nuc}}$             & $1$                 & Primary nucleation oligomer yield \\
$\alpha_{\mathrm{sec}}$             & $1$                 & Secondary nucleation oligomer yield \\
$\beta_{\mathrm{elong}}$            & $1$                 & Fibril increment from elongation \\
$\rho$                               & $0.05$              & Global network-diffusion scale \\
\bottomrule
\end{tabular}
\end{table}

The deterministic network system was integrated using a fourth-order
Runge--Kutta method to $T=300$ model-time units with
$\Delta t=0.25$. Concentrations were constrained to remain
non-negative after each numerical update. Simulation time is therefore
reported in model-time units and is not directly calibrated to patient
age or clinical years. Network-average oligomer and fibril burdens
were calculated as

\begin{equation}
\bar c_2(t)
=
\frac{1}{N}\sum_{i=1}^{N}c_{2,i}(t),
\qquad
\bar c_3(t)
=
\frac{1}{N}\sum_{i=1}^{N}c_{3,i}(t).
\label{eq:network_means}
\end{equation}

\subsection{Mutation and parameter uncertainty analyses}
\label{sec:uncertainty}

Two complementary forms of uncertainty were examined: uncertainty in
the experimentally measured mutation scores and uncertainty in the
model parameters. For each familial mutation, the reported
nucleation-score uncertainty was represented as

\begin{equation}
\widetilde{\mathrm{NS}}_{\mu}
\sim
\mathcal{N}
\left(
\mathrm{NS}_{\mu},
\sigma_{\mu}^{2}
\right),
\qquad
\widetilde f_{\mu}
=
\exp\!\left(\widetilde{\mathrm{NS}}_{\mu}\right).
\label{eq:mutation_uncertainty}
\end{equation}

WT remained fixed at $\mathrm{NS}_{\mathrm{WT}}=0$. Three alternative
mechanistic mappings were then tested: mutation effects on primary
nucleation only, secondary nucleation only, or both processes. These
experiments used the local no-diffusion model so that mutation-dependent
aggregation kinetics could be examined independently of network
topology. For each of the eight mutation conditions and each of the
three mappings, 500 Monte Carlo realizations were simulated to
$T=300$ using $\Delta t=0.10$, giving 12,000 local simulations in
total. A common WT-derived oligomer threshold,
$\theta_{\mathrm{local}}
=0.5\max_t c_2^{(\mathrm{WT})}(t)$, was used for all mutation
conditions. The principal outputs were threshold-crossing time, peak
oligomer concentration, and oligomer area under the curve,

\begin{equation}
\mathrm{AUC}_{c_2}
=
\int_0^T c_2(t)\,dt.
\label{eq:local_auc}
\end{equation}

A separate global sensitivity analysis was carried out on the full
connectome model using Latin hypercube sampling
\cite{McKay1979,Marino2008}. Ten parameters were varied simultaneously:
$k_{\mathrm{nuc}}$, $k_{\mathrm{sec}}$, $k_{\mathrm{elong}}$,
$k_{\mathrm{frag}}$, $k_{\mathrm{conv}}$, $\delta_1$, $\delta_2$,
$\delta_3$, $s$, and $\rho$. Five hundred parameter sets were sampled
independently on a logarithmic scale from $0.1$ to $10$ times each
nominal value. Sensitivity was evaluated for threshold-crossing time,
peak network-mean oligomer burden, oligomer AUC, and final spatial
spread. Parameter importance was quantified using partial rank
correlation coefficients (PRCCs), with 95\% confidence intervals
estimated from 500 nonparametric bootstrap resamples
\cite{Marino2008}.

\subsection{Network robustness and spatial propagation analyses}
\label{sec:network_robustness}

A series of robustness analyses was used to determine whether the
mutation-dependent conclusions were sensitive to the choice of seed
regions, edge-weight representation, or exact connectome topology.
Seed robustness was assessed using alternative 16-node configurations:
the 16 highest-strength candidate nodes, all 16 posterior cingulate
candidates, the 16 highest-strength precuneus candidates, and a
balanced configuration containing eight posterior cingulate and eight
precuneus nodes. Twenty additional random 16-node seed sets were drawn
from the complete pool of 36 candidate nodes. For each seed
configuration, all eight mutation conditions were simulated, and
mutation rankings for threshold-crossing time, peak oligomer burden,
and oligomer AUC were compared using Spearman rank correlation.
Regional oligomer patterns were also compared with those from the
primary 18-node seed configuration.

The effect of structural-weight representation was examined using the
original weighted connectome, a binary version of the same topology,
and a square-root transformation of the original edge weights. To test
the role of the empirical network topology itself, 20
degree-preserving randomized connectomes were generated using
connected double-edge rewiring \cite{MaslovSneppen2002}. Each null
network preserved the number of nodes and edges, the exact degree
sequence, graph connectedness, and the complete global set of
empirical edge weights, while the positions of individual edges and
weights were randomized. All null connectomes were normalized using
the same empirical normalization constant, $\kappa=782$, so that the
diffusion scale remained comparable with that of the original
connectome. Simulating all eight mutations on all 20 null networks
produced 160 randomized-network simulations. Global mutation rankings
were compared with those from the empirical connectome, while
regional patterns were compared using Spearman correlation, cosine
similarity, total-variation distance, and overlap between the 10\%
highest-burden nodes.

Spatial propagation was further examined by relating node-level
oligomer dynamics to weighted distance from the seed set. Because
larger structural weights represent stronger connections, an effective
edge length was defined as

\begin{equation}
\ell_{ij}
=
\frac{1}{
\max\!\left(A_{ij}^{\mathrm{raw}},10^{-12}\right)
}.
\label{eq:effective_distance}
\end{equation}

Weighted shortest-path distance from the nearest seed was calculated
for every node. Oligomer arrival times were then defined using 5\%,
10\%, and 20\% of each mutation's maximum node-level oligomer burden,
with the 5\% threshold used for primary reporting. Spearman
correlation was used to quantify the association between seed distance
and arrival time, and regional oligomer burden at a common WT-derived
reference time was also correlated with seed distance. These analyses
were intended as model-internal tests of connectome-constrained
propagation.

\subsection{ABC-SMC parameter recovery and posterior-predictive analysis}
\label{sec:abc}

Approximate Bayesian computation with sequential Monte Carlo
(ABC-SMC) was used to examine internal parameter recoverability
\cite{Toni2009}. No patient, imaging, or independent biological data
were used in this analysis. Instead, pseudo-observed summaries were
generated synthetically from the nominal model using known parameter
values. The resulting ABC distribution therefore represents a
synthetic parameter-recovery experiment rather than a clinically
calibrated posterior.

The inferred parameter vector was

\begin{equation}
\boldsymbol{\theta}
=
\left(
\gamma,
k_{\mathrm{sec}},
k_{\mathrm{frag}},
k_{\mathrm{conv}}
\right),
\label{eq:abc_theta}
\end{equation}

where $\gamma$ controls the strength of the mutation-score effect,

\begin{equation}
k_{\mathrm{nuc}}^{(\mu)}
=
k_{\mathrm{nuc}}^{(\mathrm{WT})}
\exp\!\left(
\gamma\,\mathrm{NS}_{\mu}
\right).
\label{eq:abc_gamma}
\end{equation}

The generating values were
$\gamma=1$, $k_{\mathrm{sec}}=10^{-3}$,
$k_{\mathrm{frag}}=10^{-4}$, and $k_{\mathrm{conv}}=10^{-1}$.
For each mutation, four summary statistics were used:
threshold-crossing time, peak oligomer concentration, oligomer AUC,
and final fibril concentration,

\begin{equation}
\mathbf s_\mu
=
\left(
t_{50,\mu},
\max_t c_{2,\mu}(t),
\mathrm{AUC}_{c_2,\mu},
c_{3,\mu}(T)
\right).
\label{eq:abc_summary}
\end{equation}

Combining these four summaries across all eight mutation conditions
produced a 32-dimensional target vector. The priors were specified in
log-parameter space, with
$0.5\leq\gamma\leq1.5$ and
$0.25\theta_0\leq\theta\leq4\theta_0$ for
$\theta\in\{k_{\mathrm{sec}},k_{\mathrm{frag}},k_{\mathrm{conv}}\}$.
A pilot sample of 750 parameter vectors was used to initialize the
tolerance schedule. Four ABC-SMC populations were then generated,
each containing 250 accepted particles, with progressively decreasing
tolerances. Acceptance rate, particle distance, and effective sample
size were recorded as convergence diagnostics.

The final weighted ABC population was subsequently propagated through
the full 540-node connectome. One hundred posterior draws were
selected using systematic resampling, and all eight mutation
conditions were simulated for each draw while the connectome, seed
set, and nominal diffusion scale were held fixed. Posterior-predictive
robustness was evaluated for aggregation timing, peak oligomer burden,
cumulative oligomer burden, mutation rankings, and regional spatial
patterns. Because the posterior was obtained from synthetic summaries
generated by the same mechanistic simulator, these analyses measure
internal robustness and do not constitute external biological or
clinical validation.

\subsection{Reaction-level stochastic model}
\label{sec:stochastic}

To test whether the deterministic mutation-dependent conclusions were
robust to intrinsic reaction fluctuations, the local three-species
aggregation model was extended using the Chemical Langevin Equation
(CLE) \cite{Gillespie2000}. The stochastic analysis was restricted to
the local reaction system; connectome diffusion remained deterministic.
It therefore evaluates reaction-level stochasticity rather than
stochastic transport between brain regions.

Let

\begin{equation}
\mathbf X(t)
=
\begin{pmatrix}
c_1(t)\\
c_2(t)\\
c_3(t)
\end{pmatrix},
\end{equation}

and let $\mathbf S$ denote the stoichiometric matrix for the nine
reaction channels. The corresponding propensity vector was

\begin{equation}
\mathbf a(\mathbf X)
=
\begin{pmatrix}
s\\
k_{\mathrm{nuc}}c_1^n\\
k_{\mathrm{sec}}c_1^mc_3^p\\
k_{\mathrm{elong}}c_1c_3\\
k_{\mathrm{conv}}c_2\\
k_{\mathrm{frag}}c_3\\
\delta_1c_1\\
\delta_2c_2\\
\delta_3c_3
\end{pmatrix}.
\label{eq:cle_propensities}
\end{equation}

The stochastic dynamics were represented as

\begin{equation}
d\mathbf X
=
\mathbf S\mathbf a(\mathbf X)\,dt
+
\frac{1}{\sqrt{\Omega}}
\mathbf S
\operatorname{diag}
\left(
\sqrt{\mathbf a(\mathbf X)}
\right)
d\mathbf W,
\label{eq:cle}
\end{equation}

where $\Omega$ controls the magnitude of intrinsic fluctuations and
$\mathbf W$ is a vector of independent Wiener processes. The CLE was
solved using the Euler--Maruyama method \cite{Higham2001}, and small
negative concentrations produced by Gaussian increments were projected
back to zero after each numerical update.

The main stochastic analysis used $T=150$, $\Delta t=0.05$, and
$\Omega=500$. Three hundred independent trajectories were generated
for each of the eight mutation conditions, giving 2400 stochastic
trajectories in total. Threshold-crossing time, peak oligomer burden,
oligomer AUC, and final fibril burden were calculated for each
realization, and stochastic mutation rankings were compared with the
corresponding deterministic ordering. Sensitivity to the CLE
system-size parameter was additionally assessed for
$\Omega\in\{100,500,2000\}$ using WT and E22G, with 200 stochastic
replicates for each value. Because $\Omega$ was not calibrated from
measured molecular copy numbers, this analysis was interpreted as a
qualitative robustness test rather than a quantitatively calibrated
model of molecular noise.

\begin{table}[htbp]
\centering
\small
\caption{Summary of uncertainty, robustness, and model-validation analyses.}
\label{tab:analysis_overview}
\renewcommand{\arraystretch}{1.20}

\begin{tabularx}{\textwidth}{
p{3.2cm}
p{5.0cm}
X
}
\toprule
\textbf{Analysis} &
\textbf{What was varied or tested} &
\textbf{Purpose} \\
\midrule

Mutation-score uncertainty &
GSE151147 nucleation scores sampled using their reported uncertainty &
Assess dependence of mutation-specific conclusions on experimental
score uncertainty \\

Mechanism uncertainty &
Mutation effect applied to primary nucleation, secondary nucleation,
or both &
Assess dependence on the assumed kinetic location of the mutation
effect \\

Global sensitivity analysis &
Ten kinetic, clearance, production, and diffusion parameters varied
using Latin hypercube sampling &
Identify the parameters most strongly associated with timing, burden,
and spatial spread \\

Seed robustness &
Alternative posterior-cingulate and precuneus seed configurations &
Assess dependence on the choice of initial seed regions \\

Connectome-weight robustness &
Weighted, binary, and square-root-transformed edge weights &
Assess dependence on the representation of structural connectivity \\

Randomized-connectome null model &
Twenty degree-preserving rewired connectomes &
Separate mutation-dependent kinetics from effects of empirical
network topology \\

Spatial propagation analysis &
Weighted distance from the seed set and node-level arrival times &
Assess whether propagation follows a connectome-distance gradient \\

ABC-SMC parameter recovery &
$\gamma$, $k_{\mathrm{sec}}$, $k_{\mathrm{frag}}$, and
$k_{\mathrm{conv}}$ &
Assess internal recoverability of selected parameters using synthetic
model-generated summaries \\

Posterior-predictive analysis &
ABC-SMC parameter uncertainty propagated through the full connectome &
Assess robustness of network-level conclusions to parameter
uncertainty \\

Chemical Langevin analysis &
Reaction-level stochastic fluctuations &
Assess robustness of deterministic mutation effects to intrinsic
reaction noise \\

\bottomrule
\end{tabularx}
\end{table}


\section{Results}
\label{sec:results}


\subsection{Mutation-specific aggregation dynamics on the empirical connectome}
\label{sec:results_dynamics}

Mutation-specific nucleation scaling produced clear differences in the
timing and cumulative burden of A$\beta$ aggregation on the empirical
540-node structural connectome. The strongest separation was observed
for oligomer threshold-crossing time. WT reached the fixed threshold at
110.67 model-time units, whereas A2V crossed later at 117.27. H6R and
A21G remained close to WT, with crossing times of 110.10 and 110.00,
respectively. Progressively earlier threshold attainment was observed
for E22K (108.07), D23N (105.75), and E22Q (102.67), while E22G showed
a markedly earlier transition at 8.80 model-time units
(Table~\ref{tab:deterministic_results}).

Peak network-mean oligomer burden varied much less across mutations
than threshold-crossing time. Peak values ranged from 0.003553 for
A2V to 0.003627 for E22G. In contrast, cumulative oligomer burden
showed a clearer mutation-dependent gradient. A2V had the lowest
oligomer AUC (0.6027), followed by WT (0.6326), whereas progressively
higher values were obtained for E22K (0.6480), D23N (0.6644), E22Q
(0.6920), and E22G (0.7225). Thus, the largest mutation-dependent
differences were expressed in aggregation timing and cumulative burden
rather than in absolute peak oligomer concentration.

\begin{figure}[htbp]
    \centering
    \includegraphics[width=0.85\textwidth]{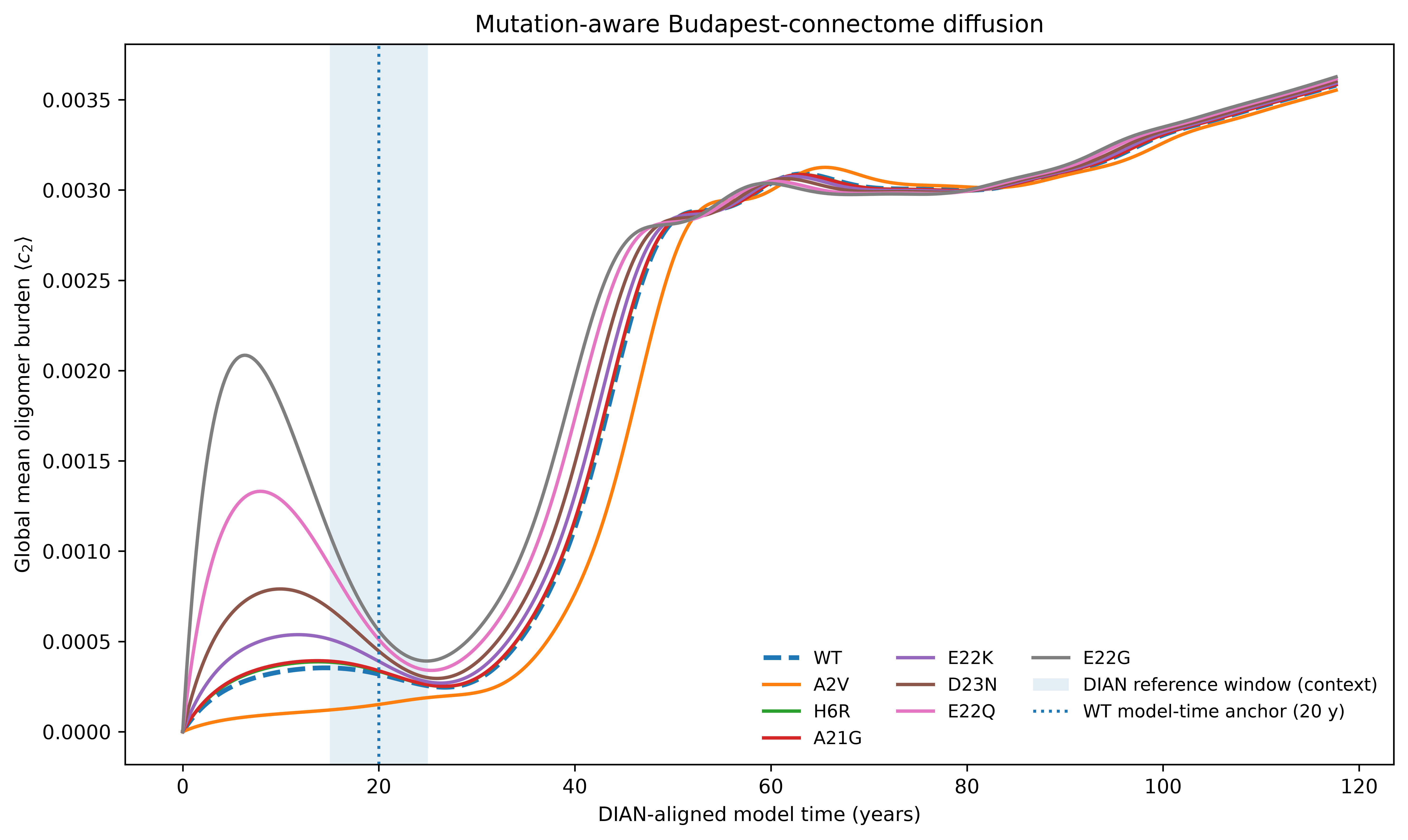}
    \caption{
    Mutation-specific oligomer dynamics on the empirical structural
    connectome. Curves show the network-mean oligomer burden for WT and
    the seven familial A$\beta$ variants. Time is expressed in
    model-time units.
    }
    \label{fig:connectome_dynamics}
\end{figure}

This ordering is consistent with experimental evidence that familial
A$\beta$ substitutions can substantially alter nucleation and
aggregation kinetics \cite{Seuma2021,Yang2018}. In particular, the
strong effect observed for E22G is concordant with experimental work
showing that the Arctic mutation promotes rapid formation of
protofibrillar A$\beta$ assemblies \cite{Nilsberth2001}. The
connectome simulations therefore retained biologically plausible
mutation-specific kinetic differences after coupling the local
aggregation model to network diffusion
(Figure~\ref{fig:connectome_dynamics}).

\begin{table}[htbp]
\centering
\small
\caption{
Mutation-specific deterministic results on the empirical structural
connectome. Threshold-crossing time is expressed in model-time units.
}
\label{tab:deterministic_results}
\renewcommand{\arraystretch}{1.15}

\begin{tabular}{lrrr}
\toprule
\textbf{Mutation} &
\textbf{$t_{50}$} &
\textbf{Peak mean $c_2$} &
\textbf{AUC of mean $c_2$} \\
\midrule
WT   & 110.67 & 0.003581 & 0.6326 \\
A2V  & 117.27 & 0.003553 & 0.6027 \\
H6R  & 110.10 & 0.003583 & 0.6358 \\
A21G & 110.00 & 0.003584 & 0.6363 \\
E22K & 108.07 & 0.003592 & 0.6480 \\
D23N & 105.75 & 0.003602 & 0.6644 \\
E22Q & 102.67 & 0.003615 & 0.6920 \\
E22G &   8.80 & 0.003627 & 0.7225 \\
\bottomrule
\end{tabular}
\end{table}

\subsection{Mutation-score uncertainty and alternative kinetic mechanisms}
\label{sec:results_mechanism}

Mutation-dependent aggregation timing remained evident when uncertainty
in the experimentally measured nucleation scores was propagated through
the local aggregation model. Under the primary-nucleation formulation
used in the principal network analysis, WT reached the local reference
threshold at 7.4 model-time units. A2V showed a markedly delayed
response: only 64\% of its 500 realizations crossed the threshold, and
the median crossing time among those realizations was 61.2 model-time
units. H6R and A21G remained comparatively close to WT, with median
crossing times of 6.3 and 6.1, respectively. In contrast, E22K, D23N,
E22Q, and E22G showed progressively earlier crossing, with median times
of 3.7, 2.2, 1.1, and 0.6 model-time units, respectively.

\begin{figure}[htbp]
    \centering
    \includegraphics[width=0.90\textwidth]{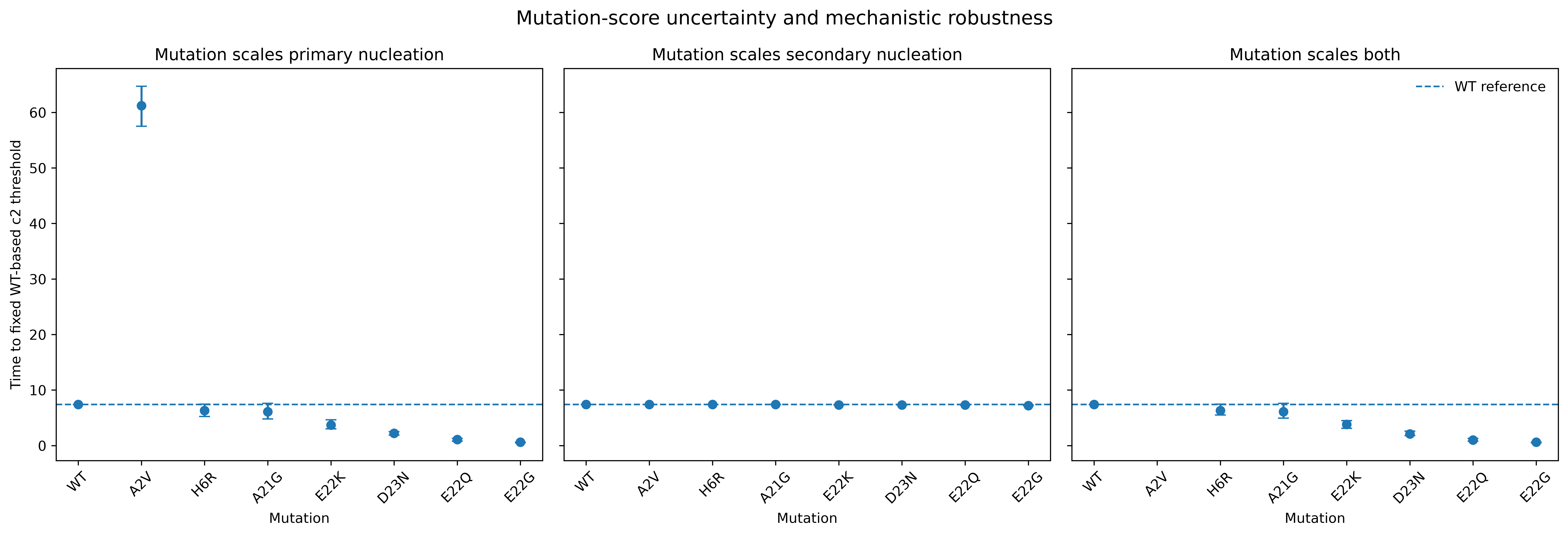}
    \caption{
    Mutation-specific threshold-crossing times under alternative
    kinetic mappings of the experimental nucleation score. Mutation
    effects were applied to primary nucleation, secondary nucleation,
    or both processes. Distributions represent uncertainty propagated
    from the experimentally measured mutation scores.
    }
    \label{fig:mechanism_uncertainty}
\end{figure}

The magnitude of these effects depended on where mutation sensitivity
was introduced into the reaction system. Applying the mutation factor
to primary nucleation produced the strongest separation in
threshold-crossing time, whereas secondary-nucleation scaling produced
a different quantitative response. When both primary and secondary
nucleation were scaled, A2V did not attain a finite median crossing
time within the simulation window, while the rapidly nucleating
variants remained clearly separated from WT
(Figure~\ref{fig:mechanism_uncertainty}). Previous kinetic experiments similarly indicate that familial
A$\beta_{42}$ substitutions can alter nucleation pathways rather than
producing a uniform change in all aggregation processes
\cite{Yang2018}. The present results do not imply that all variants act
exclusively through primary nucleation; instead, they show that the
mutation-dependent timing signal is preserved while its magnitude
depends on the mechanistic mapping used. The distinct behavior of A2V
is also consistent with experimental evidence that this variant follows
an unusual and context-dependent assembly pathway \cite{Messa2014}.

\subsection{Global parameter sensitivity identifies kinetic control of aggregation timing}
\label{sec:results_sensitivity}

\begin{figure}[htbp]
    \centering
    \includegraphics[width=0.78\textwidth]{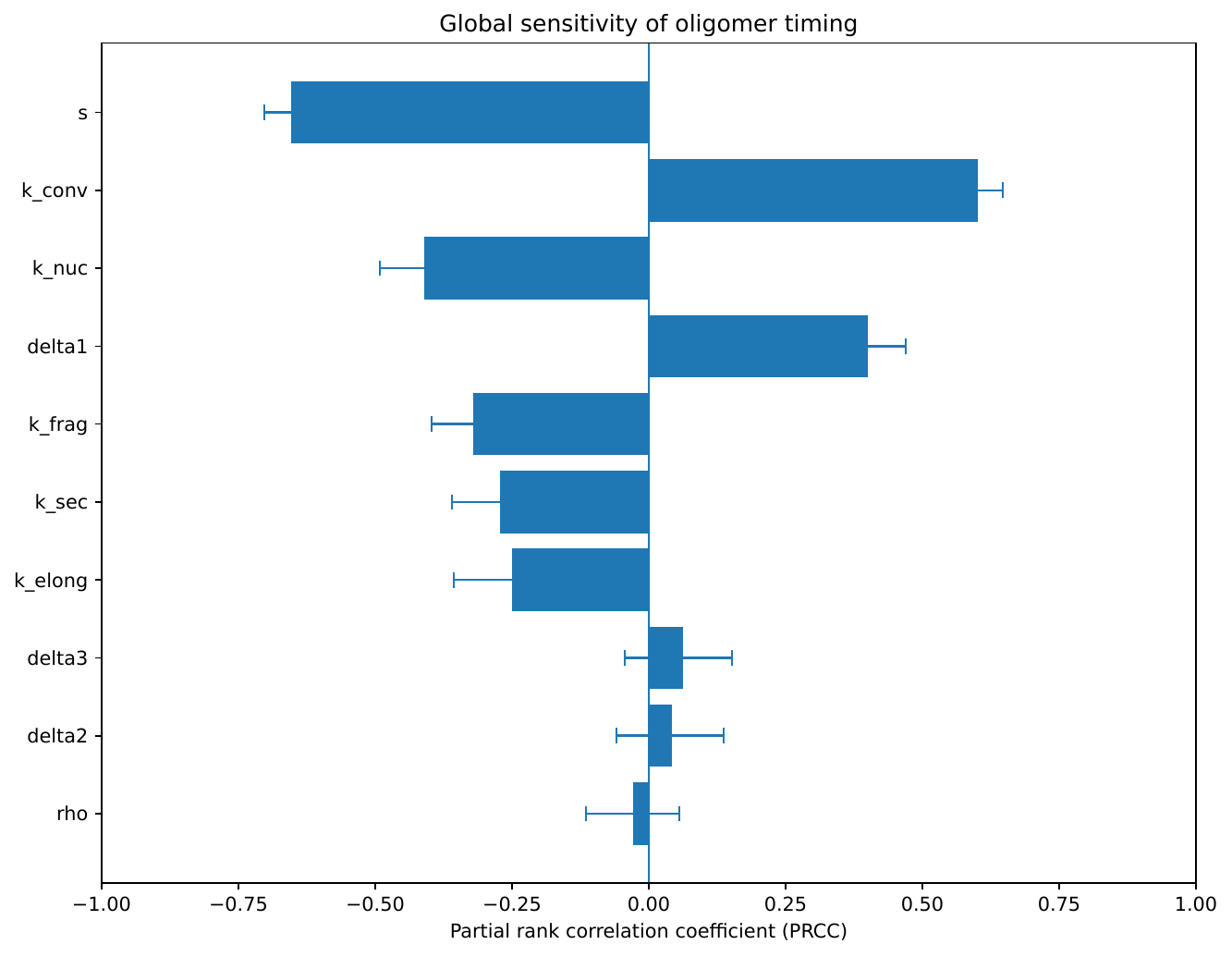}
    \caption{
    Global sensitivity of oligomer threshold-crossing time. Points
    show partial rank correlation coefficients for the ten varied model
    parameters, with 95\% bootstrap intervals. Negative coefficients
    indicate earlier crossing with increasing parameter values, whereas
    positive coefficients indicate later crossing.
    }
    \label{fig:global_sensitivity}
\end{figure}

The global sensitivity analysis demonstrated that oligomer
threshold-crossing time was primarily associated with kinetic
production, conversion, nucleation, and clearance parameters rather
than with the overall network-diffusion scale. Monomer production
showed the strongest negative association with crossing time
(PRCC $=-0.654$, 95\% bootstrap interval
$[-0.702,-0.598]$), indicating earlier threshold attainment as
production increased. Oligomer-to-fibril conversion showed a strong
positive association with crossing time
(PRCC $=0.601$, $[0.549,0.647]$), whereas primary nucleation was
associated with earlier crossing
(PRCC $=-0.411$, $[-0.492,-0.317]$). Monomer clearance was associated
with later crossing
(PRCC $=0.400$, $[0.320,0.470]$). Fragmentation
(PRCC $=-0.321$), secondary nucleation
(PRCC $=-0.273$), and elongation
(PRCC $=-0.251$) were also negatively associated with
threshold-crossing time
(Figure~\ref{fig:global_sensitivity}).

In contrast, the global diffusion parameter $\rho$ showed little
association with threshold-crossing time
(PRCC $=-0.029$, 95\% interval $[-0.115,0.055]$). Its associations
with peak oligomer burden
(PRCC $=0.020$, $[-0.067,0.109]$) and oligomer AUC
(PRCC $=0.061$, $[-0.027,0.157]$) were likewise weak. However,
$\rho$ was positively associated with the spatial-spread outcome
(PRCC $=0.317$, $[0.236,0.391]$). For cumulative oligomer burden,
monomer production (PRCC $=0.811$) and oligomer-to-fibril conversion
(PRCC $=-0.721$) were the strongest associations, whereas elongation
was the dominant parameter for final spatial spread
(PRCC $=0.878$).

These results distinguish two related but different components of the
model: local reaction kinetics primarily controlled the speed and
magnitude of network-wide aggregation, whereas connectome transport
became more relevant when the outcome explicitly measured spatial
distribution. This separation is broadly consistent with previous
connectome-based models in which local production or clearance
processes interact with structural connectivity to determine regional
A$\beta$ deposition \cite{IturriaMedina2014}.

\subsection{Mutation ordering is robust to network perturbation, whereas regional patterns depend on topology}
\label{sec:results_network}

Mutation-specific global ordering was highly stable to changes in the
initial seed configuration. Comparison of the primary 18-node seed set
with the 16-node high-strength configuration produced identical
mutation rankings for threshold-crossing time, peak oligomer burden,
and oligomer AUC
(Spearman $\rho=1$ for all three metrics). The same ordering was
retained across the structured posterior-cingulate, precuneus, and
balanced seed configurations and across the random-seed experiments.
Similarly, mutation rankings remained unchanged when the empirical
connectome was represented using its original weighted edges, binary
edges, or square-root-transformed weights.

\begin{figure}[htbp]
    \centering
    \includegraphics[width=0.82\textwidth]{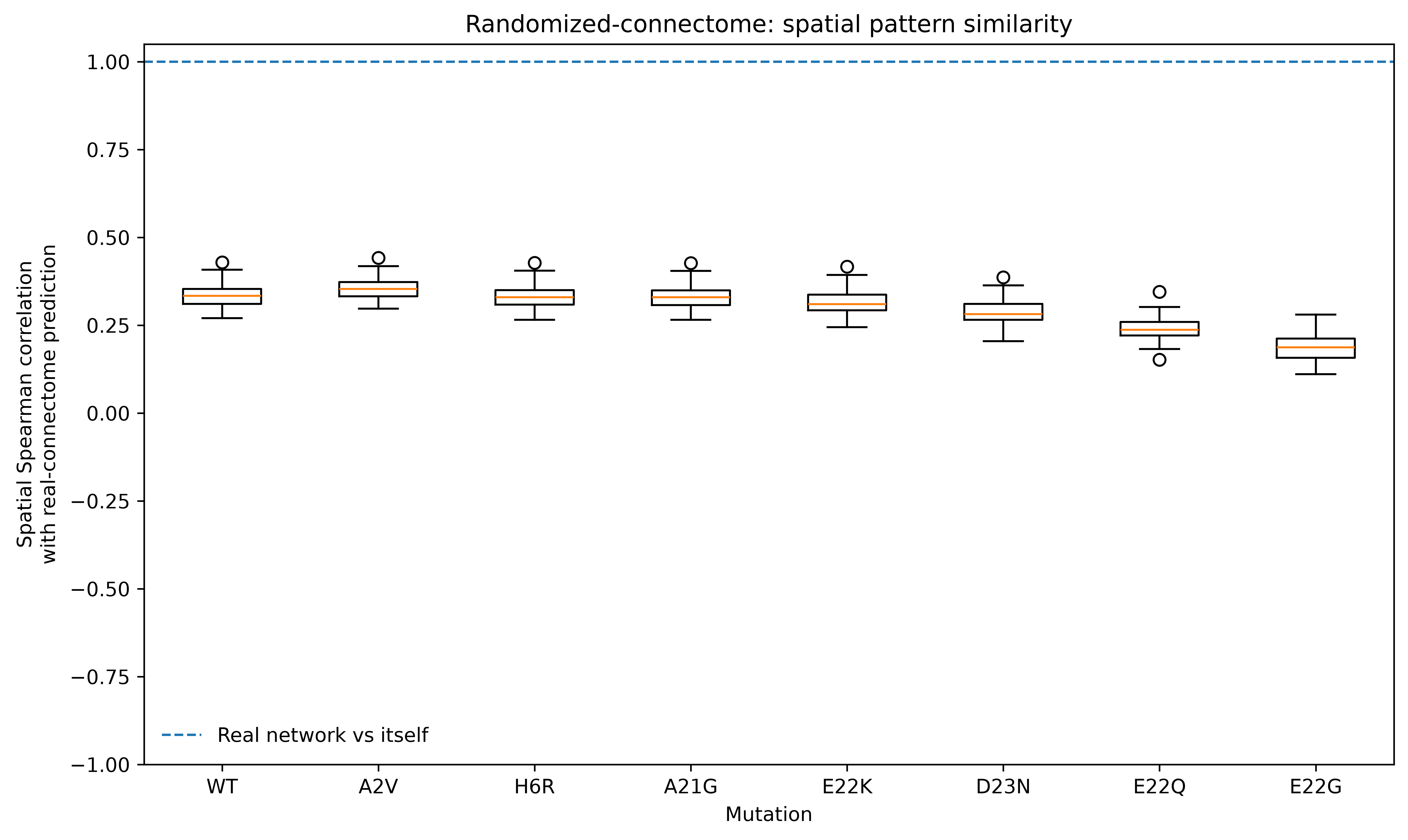}
    \caption{
    Regional oligomer-pattern similarity after degree-preserving
    connectome randomization. Boxplots show Spearman correlations
    between the regional pattern obtained on the empirical connectome
    and the corresponding patterns generated across 20 randomized
    networks for each mutation. The dashed reference line at one
    represents perfect spatial agreement.
    }
    \label{fig:randomized_connectome}
\end{figure}

The distinction between global kinetics and regional spatial
expression became clearer after randomization of the connectome
topology. All 20 degree-preserving randomized networks preserved the
mutation rankings for threshold-crossing time, peak oligomer burden,
and oligomer AUC. However, the node-level oligomer distributions were
substantially different from those produced by the empirical
connectome. Median Spearman correlations between empirical and
randomized spatial patterns were 0.333 for WT, 0.353 for A2V, 0.330
for H6R, 0.329 for A21G, 0.310 for E22K, 0.282 for D23N, 0.237 for
E22Q, and 0.187 for E22G
(Figure~\ref{fig:randomized_connectome}).

Agreement among the highest-burden regions was also limited. Median
top-10\% Jaccard overlap ranged from 0.069 for E22G to 0.125 for A2V.
Thus, extensive rewiring of the structural network left the relative
mutation ordering intact while changing the regional pattern through
which the resulting burden was expressed. This result is compatible
with network-diffusion frameworks in which structural connectivity
constrains the spatial distribution of neurodegenerative processes
\cite{raj2012network,IturriaMedina2014}.

\subsection{Oligomer propagation follows structural distance from the seed regions}
\label{sec:results_spatial}

\begin{figure}[htbp]
    \centering
    \includegraphics[width=0.82\textwidth]{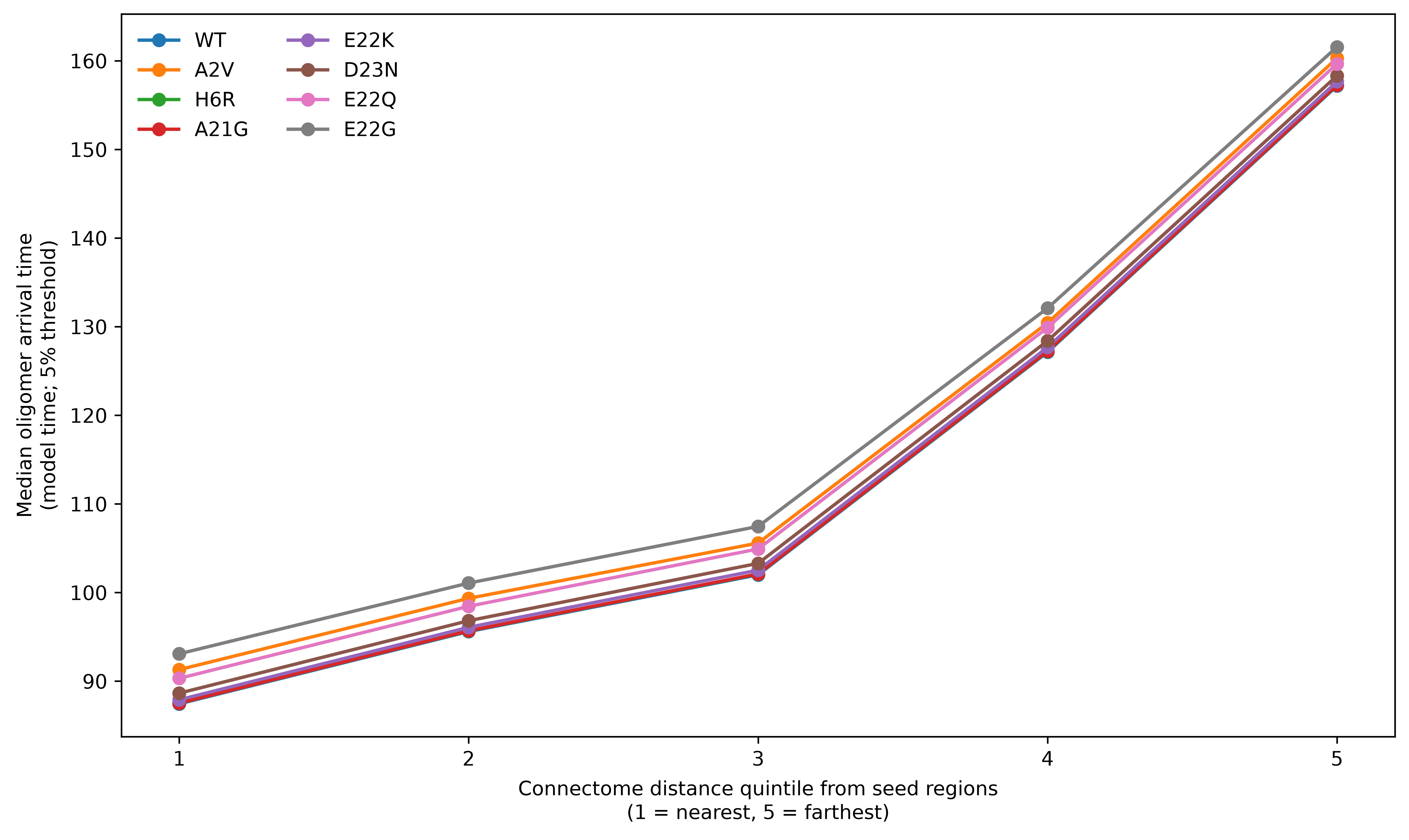}
    \caption{
    Median oligomer arrival time across weighted connectome-distance
    quintiles from the seed regions. Brain regions were divided into
    five groups according to their structural distance from the
    posterior-cingulate/\allowbreak precuneus seed set, with quintile 1 representing
    the nearest regions and quintile 5 the farthest. Arrival time was
    defined using the primary 5\% mutation-specific oligomer threshold
    and is expressed in model-time units. Across all mutation
    conditions, regions located farther from the seed set were reached
    progressively later.
    }
    \label{fig:distance_arrival}
\end{figure}

A strong spatial propagation gradient was observed between the
posterior-cingulate/\allowbreak precuneus seed set and the remainder of the
connectome. Using the primary 5\% mutation-specific oligomer threshold,
all 522 non-seed nodes crossed the threshold for every mutation. The
Spearman correlation between weighted connectome distance from the
seed set and oligomer arrival time was approximately 0.92 across all
eight mutation conditions, indicating that regions located farther from
the seed set were consistently reached later. The same pattern was
apparent when regions were grouped into five connectome-distance
quintiles: median oligomer arrival time increased progressively from
the nearest to the farthest regions
(Figure~\ref{fig:distance_arrival}).

A complementary spatial relationship was observed when oligomer burden
was evaluated at the common WT-derived reference time. The correlation
between connectome distance and regional oligomer burden was negative
for every mutation, ranging from $-0.926$ for A21G to $-0.824$ for
E22G. Thus, regions located farther from the seed set generally showed
lower oligomer burden at the same reference time. The consistency of
these relationships across mutations suggests that the spatial gradient
was primarily a property of connectome-constrained propagation rather
than a feature of any single familial variant.

This result is consistent with the connectome-based epidemic-spreading
analysis of Iturria-Medina et al., in which anatomical connectivity
constrained the regional timing and distribution of misfolded-protein
propagation \cite{IturriaMedina2014}. The use of posterior cingulate
and precuneus regions as seeds is also biologically compatible with
longitudinal amyloid-PET evidence identifying these default-mode
regions among the early cortical sites of measurable A$\beta$
accumulation \cite{Palmqvist2017}.

\subsection{Synthetic ABC-SMC recovers the generating parameter values}
\label{sec:results_abc}

The ABC-SMC experiment provided an internal test of whether selected
mechanistic parameters could be recovered from model-generated summary
statistics. All four generating parameter values fell within their
corresponding 95\% posterior intervals
(Table~\ref{tab:abc_recovery}). The mutation-scaling parameter
$\gamma$ had a posterior median of 0.950 compared with its generating
value of 1.0. The posterior medians for
$k_{\mathrm{sec}}$, $k_{\mathrm{frag}}$, and $k_{\mathrm{conv}}$
were 0.00157, 0.000063, and 0.1017, compared with generating values
of 0.001, 0.0001, and 0.1, respectively. Posterior intervals differed in width across parameters, indicating
unequal practical identifiability. In particular,
$k_{\mathrm{sec}}$ and $k_{\mathrm{frag}}$ retained broader relative
uncertainty than $\gamma$ and $k_{\mathrm{conv}}$. Correlation within
the final posterior also indicated parameter trade-offs, particularly
between $k_{\mathrm{sec}}$ and $k_{\mathrm{conv}}$.

The sequential algorithm showed progressive concentration of the
accepted particles. The tolerance decreased from 7.25 in the first
population to 2.40 in the fourth population. The final population
contained 250 accepted particles from 2724 proposals, corresponding
to an acceptance rate of 9.18\%. The final effective sample size was
240.1 and the median simulator distance was 1.88. These results support
internal recoverability of the selected parameters under the specified
simulator, priors, summary statistics, and ABC distance function
\cite{Toni2009}. They should not be interpreted as a clinically or
externally calibrated biological posterior.

\begin{table}[htbp]
\centering
\small
\caption{
ABC-SMC synthetic parameter-recovery results. The 95\% intervals are
weighted posterior percentile intervals from the final ABC-SMC
population.
}
\label{tab:abc_recovery}
\renewcommand{\arraystretch}{1.15}

\begin{tabular}{lrrrr}
\toprule
\textbf{Parameter} &
\textbf{Generating value} &
\textbf{Posterior median} &
\textbf{2.5\%} &
\textbf{97.5\%} \\
\midrule
$\gamma$            & 1.0000 & 0.9497   & 0.7332   & 1.3842 \\
$k_{\mathrm{sec}}$  & 0.0010 & 0.001567 & 0.000375 & 0.003536 \\
$k_{\mathrm{frag}}$ & 0.0001 & 0.000063 & 0.000025 & 0.000217 \\
$k_{\mathrm{conv}}$ & 0.1000 & 0.1017   & 0.0648   & 0.1653 \\
\bottomrule
\end{tabular}
\end{table}


\subsection{Timing and cumulative oligomer burden remain stable under parameter uncertainty}
\label{sec:results_posterior}

\begin{figure}[htbp]
    \centering
    \includegraphics[width=0.82\textwidth]{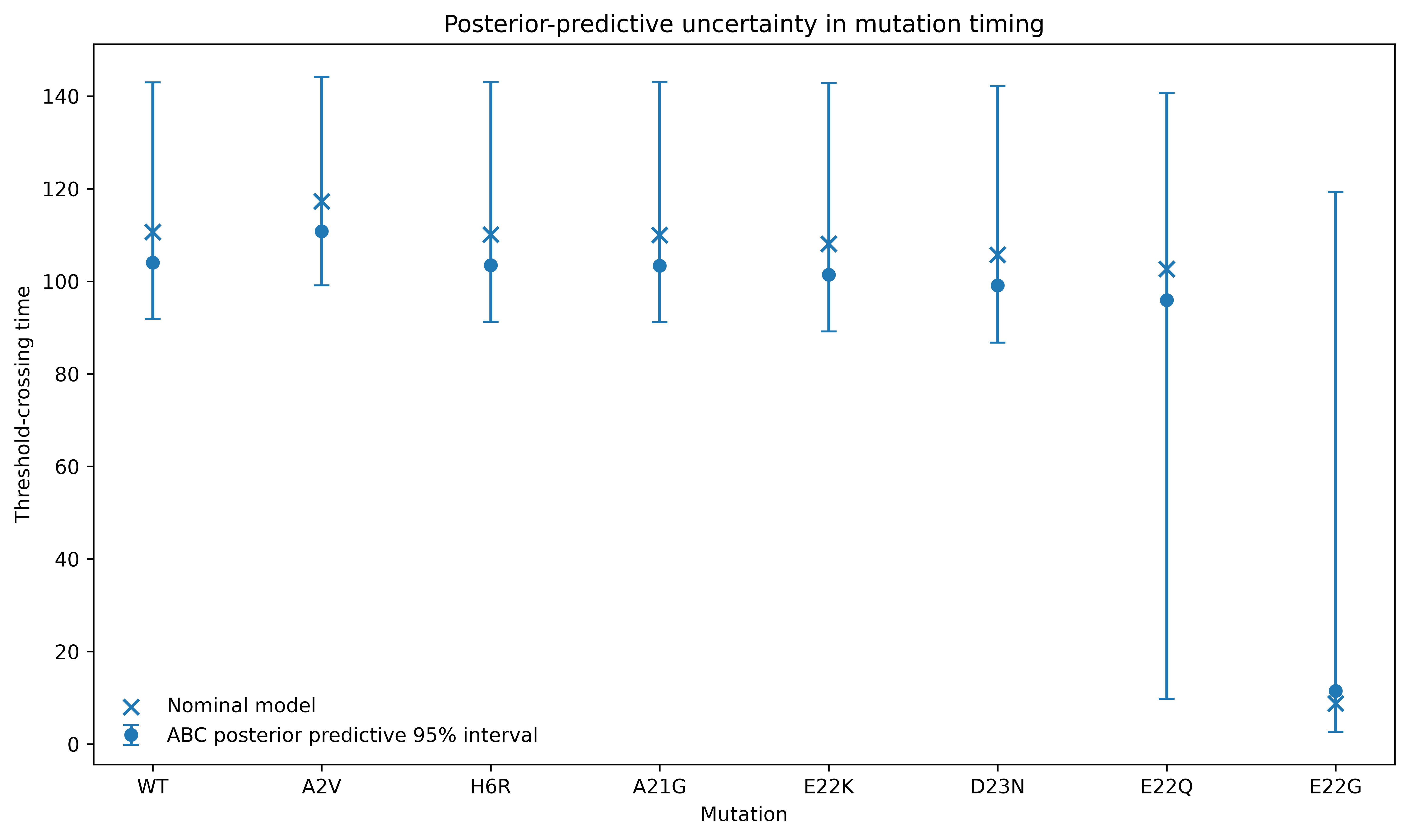}
    \caption{
    Posterior-predictive uncertainty in mutation-specific oligomer
    threshold-crossing time after propagating the ABC-SMC parameter
    distribution through the full empirical connectome. The
    posterior-predictive intervals show the variation associated with
    uncertainty in the selected kinetic parameters.
    }
    \label{fig:posterior_t50}
\end{figure}

Propagation of the ABC-SMC parameter uncertainty through the complete
540-node connectome showed that the principal conclusions for
mutation-specific timing and cumulative oligomer burden were highly
stable. Across posterior draws, the mean Spearman correlation between
the posterior-predictive and nominal mutation rankings for
threshold-crossing time was 0.990, with a median of 1.0. Exact
reproduction of the nominal timing ranking occurred in 98\% of
posterior draws (Figure~\ref{fig:posterior_t50}). The posterior probability of crossing the oligomer threshold earlier
than WT was 1\% for A2V, 99\% for H6R, A21G, E22K, and D23N, and
100\% for E22Q and E22G
(Table~\ref{tab:posterior_stochastic_results}). E22G was the fastest
mutation in every posterior draw, whereas A2V was the slowest in 99\%
of draws.

Cumulative oligomer burden was even more stable than timing. The
mutation ranking for oligomer AUC was identical to the nominal ranking
in every posterior draw. The posterior probability that cumulative
oligomer burden exceeded WT was 0\% for A2V and 100\% for H6R, A21G,
E22K, D23N, E22Q, and E22G
(Table~\ref{tab:posterior_stochastic_results}). Thus, the
mutation-specific ordering of cumulative burden was completely
preserved under the parameter uncertainty represented by the ABC-SMC
posterior. Peak oligomer amplitude was substantially less stable. Its posterior
mean mutation-rank correlation was $-0.177$, the median correlation was
$-1.0$, and the nominal peak ordering was recovered exactly in only
36\% of posterior draws. Therefore, exact peak-amplitude ranking was
more sensitive to parameter uncertainty than either threshold-crossing
time or cumulative oligomer burden. Nevertheless, all 24 nominal
scalar predictions evaluated across timing, peak burden, and AUC fell
within their corresponding 95\% posterior-predictive intervals.

Regional spatial patterns were also generally preserved under the
selected parameter uncertainty. For WT, the posterior mean spatial
Spearman correlation with the nominal regional pattern was 0.981, with
a median of 0.989. Taken together, these results indicate that
uncertainty in the selected kinetic parameters had relatively little
effect on the principal conclusions concerning aggregation timing,
cumulative oligomer burden, and broad spatial organization, while exact
peak-amplitude ordering remained less certain.

\subsection{Reaction-level stochasticity preserves ensemble mutation ordering}
\label{sec:results_stochastic}

The Chemical Langevin implementation reproduced the deterministic
reaction drift to numerical precision, with a maximum absolute
difference of $8.67\times10^{-19}$. Three hundred independent
stochastic trajectories were generated for each of the eight mutation
conditions, giving 2400 trajectories in the primary stochastic
analysis. At the ensemble level, the mutation-specific median rankings
agreed exactly with the deterministic ordering for threshold-crossing
time, peak oligomer burden, and oligomer AUC
(Spearman $\rho=1$ for all three outcomes).

Individual stochastic realizations showed greater variability than the
ensemble summaries. The median replicate-level rank correlation with
the deterministic ordering was 0.881 for threshold-crossing time and
0.929 for both peak oligomer burden and oligomer AUC. Thus, individual
stochastic trajectories could overlap even though the ordering of the
mutation-specific ensemble medians remained unchanged. The probability of crossing the oligomer threshold earlier than WT was
6.3\% for A2V, 51.3\% for H6R, 54.0\% for A21G, 73.7\% for E22K,
88.0\% for D23N, and 99.3\% for both E22Q and E22G
(Table~\ref{tab:posterior_stochastic_results}). The probabilities for
H6R and A21G were close to 50\%, indicating substantial stochastic
overlap with WT, whereas the probabilities for E22Q and E22G showed a
much clearer tendency toward earlier crossing.

Across complete stochastic realizations, E22G was strictly the fastest
mutation in 76.7\% of simulations, whereas A2V was strictly the slowest
in 90.0\%. Reaction-level stochasticity therefore introduced
variation among individual trajectories without changing the
deterministic ordering when the stochastic responses were summarized
at the ensemble level. Increasing the CLE system-size parameter reduced the discrepancy
between the stochastic ensemble mean and the corresponding
deterministic trajectory. For WT, normalized RMSE decreased from
0.205 at $\Omega=100$ to approximately 0.035 at $\Omega=500$ and
0.008 at $\Omega=2000$. A similar convergence pattern was observed
for E22G. Because $\Omega$ was not calibrated to measured molecular
copy numbers, this analysis was interpreted as a robustness test of
reaction-level fluctuations rather than as a quantitatively calibrated
estimate of biological molecular noise
\cite{Gillespie2000,Higham2001}.


\begin{table}[htbp]
\centering
\small
\caption{
Mutation-specific posterior-predictive and stochastic robustness
results. Posterior timing probability is the probability that a
mutation reached the oligomer threshold earlier than WT across
posterior-predictive simulations. Posterior AUC probability is the
probability that cumulative oligomer burden exceeded WT. CLE timing
probability is the probability that threshold crossing occurred earlier
than WT across reaction-level stochastic realizations.
}
\label{tab:posterior_stochastic_results}
\renewcommand{\arraystretch}{1.18}

\begin{tabular}{lccc}
\toprule
\textbf{Mutation} &
\textbf{Posterior} &
\textbf{Posterior} &
\textbf{CLE} \\
&
\textbf{$P(t_{50}<t_{50,\mathrm{WT}})$} &
\textbf{$P(\mathrm{AUC}>\mathrm{AUC}_{\mathrm{WT}})$} &
\textbf{$P(t_{50}<t_{50,\mathrm{WT}})$} \\
\midrule
WT   & --     & --     & --     \\
A2V  & 1\%    & 0\%    & 6.3\%  \\
H6R  & 99\%   & 100\%  & 51.3\% \\
A21G & 99\%   & 100\%  & 54.0\% \\
E22K & 99\%   & 100\%  & 73.7\% \\
D23N & 99\%   & 100\%  & 88.0\% \\
E22Q & 100\%  & 100\%  & 99.3\% \\
E22G & 100\%  & 100\%  & 99.3\% \\
\bottomrule
\end{tabular}
\end{table}


\subsection{Summary of robustness analyses}
\label{sec:results_summary}

Across the full set of analyses, mutation-dependent differences in
aggregation timing and cumulative oligomer burden were more stable than
exact peak-amplitude or regional-hotspot estimates. Mutation rankings
were unchanged under alternative seed configurations and alternative
connectome-weight representations and were preserved across all 20
degree-preserving randomized connectomes. Strong distance--arrival
relationships were retained across mutations, and posterior-predictive
analysis showed near-complete preservation of timing and AUC rankings.
Reaction-level stochasticity introduced overlap among individual
realizations but preserved the deterministic ordering of the ensemble
medians. In contrast, exact peak-amplitude ranking was sensitive to
parameter uncertainty, and detailed regional patterns changed
substantially when connectome topology was randomized.

\begin{table}[htbp]
\centering
\small
\caption{
Summary of the principal robustness results.
}
\label{tab:robustness_results}
\renewcommand{\arraystretch}{1.20}

\begin{tabularx}{\textwidth}{p{3.5cm} p{4.2cm} X}
\toprule
\textbf{Analysis} &
\textbf{Principal result} &
\textbf{Interpretation} \\
\midrule

Seed-set robustness &
Mutation-rank $\rho=1$ for timing, peak, and AUC &
Mutation ordering was insensitive to the tested seed configurations.
\\

Connectome-weight robustness &
Mutation ordering preserved under weighted, binary, and
square-root-transformed edges &
The principal kinetic ordering did not depend on one particular edge
weight representation.
\\

Randomized topology &
Mutation rankings preserved across all 20 randomized networks &
Global mutation ordering was robust to topology randomization.
\\

Randomized spatial pattern &
Median empirical--null spatial $\rho\approx0.19$--$0.35$ &
Detailed regional burden depended strongly on empirical network
topology.
\\

Spatial propagation &
Distance--arrival $\rho\approx0.92$ for all mutations &
Regions farther from the seed set were reached systematically later.
\\

Posterior prediction &
Mean timing-rank $\rho=0.990$; AUC-rank $\rho=1.000$ &
Timing and cumulative burden were highly stable to parameter
uncertainty.
\\

Posterior peak ranking &
Mean $\rho=-0.177$; exact nominal ranking in 36\% of draws &
Exact peak-amplitude ordering was parameter-sensitive.
\\

Chemical Langevin model &
Ensemble mutation-rank $\rho=1$ &
Mutation ordering persisted under reaction-level stochasticity.
\\

\bottomrule
\end{tabularx}
\end{table}

\FloatBarrier


\section{Discussion}
\label{sec:discussion}

This study developed a mutation-aware aggregation--diffusion framework to examine how experimentally informed differences in A$\beta$ nucleation can influence simulated aggregation at the scale of the structural brain network. The principal finding is that mutation-specific molecular kinetics and connectome topology played distinguishable roles in the model. Differences in the experimentally derived nucleation scores produced reproducible differences in aggregation timing and cumulative oligomer burden even though all mutation conditions were simulated on the same structural connectome and with the same primary seed configuration. E22G showed the earliest threshold crossing and the largest cumulative oligomer burden, E22Q and D23N showed intermediate-fast behavior, E22K produced a smaller acceleration relative to WT, H6R and A21G remained comparatively close to WT, and A2V showed the slowest deterministic trajectory under the selected mutation-to-rate mapping. Importantly, the largest mutation-dependent differences were observed in timing and cumulative burden rather than in absolute peak oligomer concentration, suggesting that the effect of the mutation-specific parameterization was expressed more strongly through when aggregation developed and how much burden accumulated over time than through a large separation in maximum network-mean oligomer level. This interpretation is consistent with the experimental basis of the model. Seuma et al. used deep mutational scanning to show that sequence variation within A$\beta_{42}$ can strongly alter fibril nucleation and that dominant familial Alzheimer disease mutations within the peptide tend to increase nucleation \cite{Seuma2021}. Experimental kinetic work has also demonstrated that familial substitutions near residues 22 and 23, including E22K, E22G, E22Q, and D23N, alter A$\beta_{42}$ aggregation through changes in nucleation-dependent processes \cite{Yang2018}. The particularly strong effect of E22G in the present simulations is biologically compatible with the original characterization of the Arctic APP mutation by Nilsberth et al., who reported substantially enhanced protofibril formation relative to WT A$\beta$ \cite{Nilsberth2001}. The present simulations should not, however, be interpreted as a ranking of clinical severity among familial mutations. They instead show the consequences of applying experimentally informed differences in nucleation phenotype to a common mechanistic and anatomical system.

The behavior of A2V illustrates why this distinction between a model-derived kinetic effect and a clinical interpretation is important. The APP A673V mutation corresponding to A2V in the A$\beta$ sequence has a complex biological phenotype: homozygous carriers have been associated with early-onset Alzheimer-type disease, whereas mixtures of mutant and WT peptide can alter aggregation differently from the mutant peptide alone \cite{DiFede2009}. Messa et al. subsequently demonstrated that A2V follows a distinctive molecular-assembly pathway and that its aggregation behavior depends on peptide composition and intermolecular interactions \cite{Messa2014}. The delayed trajectory in the present model is therefore best interpreted as the consequence of the particular experimental nucleation phenotype and mapping used here rather than as a general statement about the clinical effect of A2V. The mechanism-uncertainty analysis reinforces this point more broadly. Mutation-dependent differences persisted when the experimental score was mapped to alternative reaction pathways, but the magnitude of the separation changed depending on whether mutation sensitivity was assigned to primary nucleation, secondary nucleation, or both. This is biologically reasonable because amyloid formation is governed by several coupled microscopic processes rather than by a single rate constant. Primary nucleation establishes new aggregates, whereas secondary nucleation, elongation, fragmentation, and conversion can subsequently amplify or redistribute aggregate burden \cite{Cohen2013,Knowles2014}. Experimental work by Cohen et al. demonstrated that fibril surfaces can catalyze the formation of new A$\beta_{42}$ aggregates through secondary nucleation \cite{Cohen2013}, while broader kinetic analyses have emphasized that several microscopic mechanisms can generate superficially similar macroscopic aggregation curves \cite{Knowles2014}. The mechanism analysis in the present study therefore does not establish that familial mutations act exclusively through primary nucleation; rather, it shows that experimentally observed mutation differences can remain influential at the network scale while the quantitative magnitude of that influence depends on the biochemical pathway through which they are represented.

The global sensitivity analysis further supports a distinction between the local kinetic processes controlling network-wide aggregation and the structural factors controlling its spatial expression. Monomer production, oligomer-to-fibril conversion, primary nucleation, monomer clearance, fragmentation, secondary nucleation, and elongation all influenced threshold-crossing time to different degrees, whereas the global diffusion parameter $\rho$ had little association with threshold timing, peak network-mean oligomer burden, or oligomer AUC. Primary nucleation nevertheless remained an important determinant of timing, with increasing $k_{\mathrm{nuc}}$ associated with earlier threshold crossing, which is consistent with the role of nucleation as an early rate-limiting step in amyloid assembly \cite{Knowles2014,Seuma2021}. At the same time, the comparatively strong effects of monomer production and conversion demonstrate that mutation-specific nucleation cannot be considered in isolation from the rest of the reaction system. Once aggregates are present, the balance among production, conversion, secondary nucleation, fragmentation, elongation, and clearance determines how rapidly oligomer burden accumulates and whether material remains in the oligomeric pool or progresses toward fibrillar states \cite{Cohen2013,Knowles2014}. A notable result was that diffusion became more influential when the outcome explicitly measured spatial spread, despite its weak association with the global timing and burden metrics. This pattern provides a mechanistic interpretation of the overall findings: reaction kinetics largely determine the global pace and cumulative amount of aggregation in this model, whereas interregional transport becomes increasingly important when the question concerns how that burden is distributed through the brain. This separation is also consistent with previous multiscale approaches in which local production or clearance processes interact with structural pathways to determine large-scale protein distribution \cite{IturriaMedina2014}.

The network-robustness experiments provide additional evidence for this kinetic--topological separation. Mutation rankings for timing, peak burden, and cumulative burden were preserved under alternative posterior-cingulate and precuneus seed configurations and under weighted, binary, and square-root-transformed representations of the connectome. More importantly, the same global mutation ordering was retained across all 20 degree-preserving randomized networks even though the detailed regional patterns were substantially altered. Median correlations between empirical-connectome and randomized-connectome spatial patterns were only approximately 0.19--0.35, and overlap among the highest-burden regions was low. These results indicate that the mutation-dependent kinetic signal was sufficiently strong to survive extensive changes in network organization, whereas the identity of the regions carrying the greatest burden depended much more strongly on the empirical topology. This behavior is conceptually consistent with connectome-based models of neurodegeneration. Raj et al. demonstrated that diffusion over anatomical connectivity can generate characteristic spatial patterns of neurodegenerative disease \cite{raj2012network}, while Iturria-Medina et al. developed an epidemic-spreading model in which structural connectivity, pathological-agent propagation, and clearance jointly shaped regional misfolded-protein distributions \cite{IturriaMedina2014}. The present study adds a different layer to this class of models by allowing the local molecular kinetics themselves to differ between familial A$\beta$ variants while holding the anatomical scaffold constant. The strong relationship between structural distance and simulated arrival time provides further support for this interpretation. Across all mutations, the distance--arrival Spearman correlation was approximately 0.92, and median arrival time increased progressively from the nearest to the farthest connectome-distance quintile. This result closely parallels the general prediction of network-spreading models that anatomically more distant regions are reached later when transport is constrained by structural pathways \cite{raj2012network,IturriaMedina2014}. The use of posterior cingulate and precuneus regions as principal seeds is also biologically plausible because longitudinal amyloid-PET studies have identified the precuneus and posterior cingulate among regions showing particularly early A$\beta$ accumulation in preclinical disease \cite{Palmqvist2017}. Nevertheless, the present analysis does not establish that these regions are the unique biological origin of familial amyloid pathology; the seed-robustness results instead show that the major mutation ordering was not dependent on a single exact seed configuration.

The synthetic ABC-SMC and posterior-predictive analyses provide a complementary assessment of uncertainty that was not available from the deterministic simulations alone. In the synthetic recovery experiment, the generating values of $\gamma$, $k_{\mathrm{sec}}$, $k_{\mathrm{frag}}$, and $k_{\mathrm{conv}}$ all fell within their corresponding 95\% posterior intervals, showing that the selected summary statistics contained enough information to recover the known generating values under the assumed simulator and priors. The posterior was not equally concentrated for all parameters, however, and substantial correlations between parameters, particularly between $k_{\mathrm{sec}}$ and $k_{\mathrm{conv}}$, indicated compensatory relationships in which different combinations of reaction rates could generate similar summary behavior. This is an important feature of mechanistic models because agreement with an observed trajectory does not necessarily imply that every microscopic parameter is uniquely identified. The ABC-SMC analysis is an internal parameter-recovery and practical-identifiability experiment rather than biological calibration; the pseudo-observed summaries were generated from the same model used for inference, and no patient biomarker data were used to construct the posterior \cite{Toni2009}. When the resulting parameter uncertainty was propagated through the complete connectome, the major conclusions for threshold timing and cumulative oligomer burden were remarkably stable. The mean rank correlation for threshold timing was 0.990, exact timing order was recovered in 98\% of posterior draws, and the oligomer-AUC ranking was preserved in every draw. In contrast, exact peak-amplitude ranking was unstable under parameter uncertainty, demonstrating that not every model output was equally robust. This difference is scientifically useful rather than problematic: it identifies timing and cumulative burden as more defensible mutation-level outputs of the present framework, while suggesting that fine distinctions in peak oligomer amplitude should not be emphasized without stronger parameter constraints. The posterior-predictive analysis therefore narrows the principal interpretation of the model to those quantities that remained stable when uncertainty in the kinetic parameters was explicitly propagated.

Reaction-level stochasticity addressed a different source of uncertainty. The Chemical Langevin formulation introduces fluctuations associated with individual reaction channels rather than changing the underlying kinetic parameter values, and therefore asks whether the deterministic mutation differences persist when the aggregation reactions themselves are noisy. At the ensemble level, the deterministic ordering was preserved for threshold-crossing time, peak oligomer burden, and oligomer AUC, but individual stochastic trajectories showed considerably greater overlap. Near-WT mutations such as H6R and A21G crossed earlier than WT in only about half of individual stochastic realizations, whereas E22Q and E22G crossed earlier in almost all realizations. E22G was the strictly fastest condition in 76.7\% rather than 100\% of individual stochastic simulations, demonstrating that the stochastic analysis does not imply an invariant ordering in every realization. This distinction between ensemble behavior and individual trajectories is important when interpreting stochastic mechanistic models. The Chemical Langevin equation is an approximation to reaction-level stochastic dynamics that becomes increasingly consistent with deterministic behavior as the effective system size increases \cite{Gillespie2000,Higham2001}. The observed convergence toward the deterministic trajectories as $\Omega$ increased is therefore expected and provides an internal numerical check on the stochastic formulation. However, $\Omega$ was not calibrated to molecular copy numbers in brain tissue, so the widths of the stochastic distributions should not be interpreted as estimates of patient-to-patient biological variability. Instead, the CLE analysis demonstrates that the principal mutation-dependent tendencies are not artifacts of a purely deterministic solver while simultaneously showing that stochastic fluctuations can blur distinctions between mutations whose deterministic trajectories are close to WT.

Several limitations define the scope of these findings. The mutation-specific
nucleation scores were converted into kinetic-rate multipliers through an
exponential mapping, which is a modelling assumption rather than a direct
estimate of microscopic rate constants. The three-state
monomer--oligomer--fibril system is intentionally coarse grained and does not
capture the full complexity of A$\beta$ aggregation \cite{Knowles2014}. In
addition, the Budapest Reference Connectome represents a healthy consensus
network rather than disease- or mutation-specific anatomy
\cite{szalkai2017budapest}, and important processes such as vascular amyloid
deposition, tau pathology, neuroinflammation, and region-specific clearance
are not included. Simulation time is expressed only in model-time units and
should not be interpreted as clinical years or age at onset. Finally, the
ABC-SMC analysis used synthetic pseudo-observations and therefore demonstrates
internal parameter recoverability rather than external validation. Future work
should focus on calibration against mutation-specific \textit{in vitro}
kinetics and validation using longitudinal amyloid-PET or fluid-biomarker data.
Within these limits, the model supports the conclusion that mutation-dependent
kinetics primarily influence the relative timing and cumulative burden of
aggregation, whereas connectome structure mainly shapes its spatial
distribution.

\section{Conclusion}
\label{sec:conclusion}

This study developed a mutation-aware mechanistic framework for examining how familial
A$\beta$ variants may influence simulated aggregation and propagation across a structural
brain network. Experimentally derived mutation-specific nucleation phenotypes were
incorporated into a coarse-grained Smoluchowski-type aggregation--fragmentation model
and coupled across the Budapest Reference Connectome through graph diffusion. Within
this framework, mutation-dependent differences in local aggregation kinetics were
sufficient to generate distinct network-level trajectories even when the underlying
connectome and principal seed configuration were held constant. The most consistent
differences were observed in aggregation timing and cumulative oligomer burden. E22G
showed the earliest propagation and greatest cumulative oligomer burden, E22Q and D23N
showed intermediate-fast dynamics, H6R and A21G remained comparatively close to WT,
and A2V showed delayed dynamics under the selected mutation-to-rate mapping. These
results should be interpreted as consequences of the experimentally informed kinetic
parameterization rather than as a ranking of clinical mutation severity.

A major finding of the study was that molecular kinetics and structural network topology
played related but distinguishable roles. Global sensitivity analysis showed that
aggregation timing and cumulative burden were influenced mainly by kinetic parameters,
including monomer production, nucleation, conversion, fragmentation, elongation, and
clearance, whereas the overall diffusion scale contributed more strongly when spatial
spread was considered. Consistent with this distinction, mutation rankings were
preserved across alternative seed configurations, alternative connectome-weight
representations, and all 20 degree-preserving randomized connectomes, while the detailed
regional oligomer patterns changed substantially after network randomization. The strong
positive relationship between weighted connectome distance and oligomer arrival time
further showed that structural organization constrained the spatial progression of the
simulated burden: regions located farther from the posterior-cingulate/\allowbreak precuneus seed
set were systematically reached later.

The uncertainty and stochastic analyses further clarified which model conclusions were
most robust. The synthetic ABC-SMC experiment recovered the known generating values
of the selected parameters within their 95\% posterior intervals, while also revealing
differences in practical identifiability and correlations among kinetic parameters.
Propagation of this parameter uncertainty through the complete connectome showed that
mutation-specific timing and cumulative oligomer burden were highly stable: the mean
posterior-predictive rank correlation for timing was 0.990 and the oligomer-AUC ranking
was preserved in every posterior draw. In contrast, exact peak-amplitude ordering was
substantially more sensitive to parameter uncertainty, indicating that peak burden should
be interpreted more cautiously than timing or cumulative burden. Reaction-level
stochastic simulations using the Chemical Langevin equation produced a complementary
result. Individual trajectories showed considerable variability and overlap, particularly
for mutations close to WT, but the mutation ordering of the ensemble medians remained
identical to the deterministic ordering for threshold-crossing time, peak oligomer burden,
and oligomer AUC.

Taken together, the analyses support a specific interpretation of the model: familial
mutation-dependent aggregation kinetics primarily determine the relative timing and
cumulative magnitude of simulated A$\beta$ burden, whereas the structural connectome
primarily shapes the spatial route through which that burden is expressed. This conclusion
was supported across deterministic simulations, mutation-score uncertainty, global
sensitivity analysis, seed and edge-weight perturbations, randomized-connectome
experiments, spatial propagation analysis, synthetic ABC-SMC recovery,
posterior-predictive simulations, and reaction-level stochastic modelling. At the same
time, the framework remains a mechanistic proof of concept rather than a clinically
calibrated predictive model. Simulation time is expressed in model-time units, the
mutation scores are relative nucleation phenotypes rather than measured microscopic rate
constants, the connectome represents a healthy consensus network, and the model does
not incorporate tau pathology, neuroinflammation, vascular amyloid deposition,
region-specific clearance, neuronal loss, or patient-specific biomarker trajectories.
Future extensions should therefore focus on mutation-specific kinetic calibration,
external validation using longitudinal amyloid-PET or fluid-biomarker data, incorporation
of regionally varying clearance and additional pathological processes, and evaluation on
individual or disease-specific structural networks. Such developments would allow the
present molecular-to-network framework to move beyond internal mechanistic robustness
toward stronger biological calibration and eventual subject-specific modelling.

\section*{Acknowledgements}
The authors gratefully acknowledge the institutional and research support provided by Prof. Hina Shaheen, which significantly contributed to the successful completion of this study.

\section*{Data and Code Availability}

The datasets and code supporting the findings of this study are available in the GitHub repository
\href{https://github.com/Alamgir9/Familial-Mutations-data-repository}{Familial-Mutations-data-repository}.

\section*{Funding}
The research did not receive any specific grant from funding agencies in the public, commercial or not-for-profit sectors.

\FloatBarrier

\bibliographystyle{unsrtnat}
\bibliography{reference}
\end{document}